\documentclass{article}

\pdfoutput=1
\usepackage{lscape}
\usepackage{enumerate}
\usepackage{amsfonts}
\usepackage{amsmath}
\usepackage{comment}
\usepackage{textcomp}
\usepackage{amsthm, tikz}
\usepackage{amssymb}
\usepackage{color}
\usepackage{graphicx}
\usepackage{bbm}
\usepackage[matrix,arrow,curve]{xy}
\usepackage{array}

\def\be{\begin{eqnarray}}
\def\ee{\end{eqnarray}}
\def\nn{\nonumber}

\usepackage{tikz}
\usepackage{braids}

\hoffset= - 1.0in         

\begin{document}

\title{\vspace{-.5cm}{\bf {Checks of  integrality properties\\ in topological strings }
  }\vspace{.2cm}}
\author{{\bf A. Mironov$^{a,b,c}$}, \ {\bf A. Morozov$^{b,c}$}, \ {\bf An. Morozov$^{b,c,d}$}, \\   {\bf P. Ramadevi$^{e}$}, \ {\bf Vivek  Kumar Singh$^{e}$},\
{\bf A. Sleptsov$^{b,c,d}$}
}
\date{ }

\maketitle

\vspace{-5.5cm}

\begin{center}
\hfill FIAN/TD-19/16\\
\hfill IITP/TH-14/16\\
\hfill ITEP/TH-20/16\\
\end{center}

\vspace{4.2cm}

\begin{center}

$^a$ {\small {\it Lebedev Physics Institute, Moscow 119991, Russia}}\\
$^b$ {\small {\it ITEP, Moscow 117218, Russia}}\\
$^c$ {\small {\it Institute for Information Transmission Problems, Moscow 127994, Russia}}\\
$^d$ {\small {\it Laboratory of Quantum Topology, Chelyabinsk State University, Chelyabinsk 454001, Russia }}\\
$^e$ {\small {\it Department of Physics, Indian Institute of Technology Bombay, Mumbai 400076, India}}
\end{center}

\vspace{1cm}

\begin{abstract}
Tests of the integrality properties of a scalar operator in topological strings  on a resolved conifold background or orientifold
of conifold backgrounds have been performed for arborescent knots and some non-arborescent  knots.
The recent results on  polynomials for those  knots colored by $SU(N)$  and $SO(N)$ adjoint representations
\cite{MMuniv} are useful to verify Marino's integrality conjecture up to two boxes in the Young diagram. In this paper, we review the salient
aspects of the integrality properties and tabulate explicitly for an arborescent knot and a link. In our knotebook website, we have put these results
for over 100 prime knots available in Rolfsen table and some links. The first application of the obtained results, an observation of the Gaussian distribution of the LMOV invariants is also reported.
\end{abstract}

\vspace{.1cm}

\section{Introduction}

Topological string duality conjectures put forth by Gopakumar-Vafa \cite{GV} and  Ooguri-Vafa \cite{OV} relates $U(N)$ Chern-Simons theory on $S^3$ to topological string theory on resolved conifold.
This has led to rewrite suitable combinations of Chern-Simons knot polynomials \cite{CS,knotpols,Con} as reformulated invariants possessing integrality structures \cite{LM1,RS,LMV,LM2,MV} famously known as LMOV condition (see also \cite{Zhu} for the latest development). These integer invariants count the
BPS states(spectra of M2 branes ending on M5 branes in M-theory compactified on the conifold \cite{GV}). These integers  determine the oriented  topological string amplitudes.
The challenge to obtain the integers needs polynomial form of unreduced colored HOMFLY-PT for any knot/link.  In fact, a recent breakthrough \cite{NRZ13}-\cite{MMMSsu},\cite{MMuniv} enabled evaluation of colored HOMFLY-PT polynomials. Our knotebook website \cite{knotebook} which gets updated periodically gives the list of knots for which polynomials are obtained.
Thus we can indirectly determine the BPS integers for such knots and thus verify the integrality structures within topological string duality context.

The LMOV \cite{OV,MV} integrality condition is much stronger than the integrality of colored HOMFLY-PT and Kauffman knot polynomials.
That is,   in suitable variables ($q = \exp \left(\frac{2\pi i}{k+N}\right)$
and $A = q^N$), the expectation values of Wilson loop operators
\be
P^{\cal L}_R(A,q)\ = \ \left< {\rm Tr}_R\  T\exp\left(\oint_{\cal L} {\cal A}\right)\right>
\label{WLav}
\ee
in $SU(N)$(colored HOMFLY-PT) or $SO(N+1)$ (colored Kauffman) Chern-Simons theories are Laurent polynomials with integer coefficients.
There was no topological arguments to justify these integers. The categorification technique introduced
by Khovanov-Rozansky \cite{KhR}  interprets these integers as dimension of doubly graded vector space.
It is still a challenging question to find the connection of such a categorification approach and the
conventional Reshetikhin-Turaev (RT) formalism \cite{RT}-\cite{RTmod2}. In refs.
\cite{WittenMorse,KW,Wittenlast}, these integers in Jones polynomials are interpreted as counting  solutions of Hitchin equation in a
four dimensional gauge theory for a given instanton number.

There is an elegant way of writing expectation value of Ooguri-Vafa scalar operator for knots in topological strings \cite{OV}
using  plethystic exponential of a spectrum generating function (SGF) known as index.
Technically, if a Hilbert space has a SGF
\be
Ind_{H}(t) = \sum_i \tilde N_i t^i~,
\ee
then,  the Fock space has a single state (vacuum) at the zeroth level, $M_1=\tilde N_1$ states at the
first level, $M_2 = \tilde N_2 + \frac{1}{2}\tilde N_1 (\tilde N_1+1)$ states at the second level,
$M_3 = \tilde N_3 + \tilde N_2\tilde N_1 + \frac{1}{6}\tilde N_1 (\tilde N_1+1)(\tilde N_1+1)$
at the third, and so on, and the SGF in the Fock space is the plethystic exponential of ``the free energy'' (of SGF) in the Hilbert space:
\be
Ind_F(t) = \sum_I M_I t^I = \prod_i \frac{1}{(1-t^i)^{\tilde N_i}}
= \exp \sum_{d=1} \frac{Ind_H(t^d)}{d}
\ee
Thus, if some quantity is supposed to have an interpretation as an SGF,
its plethystic logarithm should also resemble an SGF: possess integrality properties.
The original Ooguri-Gopakumar-Vafa conjecture \cite{GV,OV} for knot polynomials reflected the old belief
that they are actually {\it characters}, and the plethysm operation is well known to
act naturally on the characters \cite{Littlewood} (physically the plethysm operation in the conjecture is related with the Schwinger mechanism of brane creation \cite{GV}).
For example, the generating function of quantum dimensions
(in knot theory these are {\it unreduced} HOMFLY polynomials for the unknot) is
\be
Z_{OV}^{unknot}\{A,q\,|\,\bar p\} = \sum_R {\rm dim}_{_R}(A,q) \cdot  \chi_R\{\bar p\} =
\sum_R \chi_R(p^*) \chi_R(p) = \exp\left(\sum_d \frac{p_d^*\bar p_d}{d} \right) = \nn \\
= \exp\left(\sum_d \frac{1}{d} \widehat{Ad}_d (p_1^*)\,\widehat{Ad}_d (\bar p_1)\right)
= \exp \left(\sum_d \frac{p_1^*(A^d,q^d)}{d}\,\widehat{Ad}_d (\bar p_1)\right)~,
\label{ZOVunknot}
\ee
where the sum runs over all Young diagrams $R$ (``colors"),
$\chi_R$ are the Schur functions of time variables $\bar p_k$
and the Adams (plethysm) operation $\ \widehat{Ad}_d: \ p_k\longrightarrow p_{kd}$
acts at the {\it topological locus} \cite{MMMkn12,DMMSS} by raising the power of parameters:
\be
\widehat{Ad}_d(p_k^*) = p^*_{kd} = \frac{A^{kd}-A^{-kd}}{q^{kd}-q^{kd}} =
p_k^*(A^d,q^d)~.
\ee
The plethystic logarithm of $Z_{OV}^{unknot}$ is therefore $p_1^* = \frac{A-A^{-1}}{q-q^{-1}}$
and it possesses the integrality property:
\be
(q-q^{-1}) \cdot p_1^* = A- A^{-1}
\ee
is a Laurent polynomial in variables $A$ and $q$ with integer coefficients for any knot. In this case of unknot, it is actually independent of $q$.  The claim is that such an integrality property is true for plethystic logarithms for
the Ooguri-Vafa generating functions of colored HOMFLY for {\it all} knots.
In particular, it suggests that free energies have only the {\it first order} poles in the
Planck constant $\hbar = \log q$ which is expected for the partition function but
not so obvious for {\it unreduced} knot polynomials (
$P_R \sim \hbar^{-|R|}$).

As follows from (\ref{ZOVunknot}), the plethystic transform exactly compensates the deviation,
 of the  dimension $d_R$ from $\frac{d_1^{|R|}}{|R|!}$ (both classical and quantum dimensions).
Further the deviation of the HOMFLY polynomial of general knots  from the quantum dimension, i.e. non-classical nature of {\it cabling}, is measured by the Ooguri-Vafa polynomials $f_R$(reformulated invariants).
Basically, these are homogeneous polylinear combinations
\be
f_R = H_R + \sum_I c_I \prod_{i\in I} \widehat{Ad}_{n_i}\!\big( H_{R_i}\big)
\ee
with $\sum_i n_i|R_i|=|R|$ and all $|R_i|<|R|$, which vanish if all HOMFLY polynomials are substituted by the dimensions,
 $H_R\longrightarrow d_R$, e.g. $f_{[2]} = H_{[2]} - \frac{1}{2}H_{[1]}^2-\frac{1}{2}\widehat{Ad}_2(H_1)$.
This property defines them up to triangular transforms,
and they automatically have only the first-order poles in $\{q\}(= q-q^{-1})$.

While similarity between knot polynomials and characters of an infinite-dimensional algebra is still a plausible conjecture (see some examples in \cite{MMSh}),
the fact is that these {\it averages}(\ref{WLav}) for arbitrary representations involves character decomposition.
Further localization ideas {\it a la} \cite{DH,PMN,Wittenloc,Carlson,Pestun}
can convert these averages  into a finite-dimensional matrix model integral satisfying the
AMM/EO topological recursion \cite{AMM/EO}.
This has been achieved for the torus knots \cite{BEMT}.  There are still difficulties
implementing AMM/EO topological recursion  for twist knots \cite{AMMMmamotwist} but there is some  evidence of applicability of AMM/EO recursion to  the non-torus knot $4_1$ \cite{DFM}.

Motivated by the  t'Hooft large $N$  {\it  genus expansion} (closed string partition function) for
free energies in gauge theories, we do expect genus expansion in $\hbar$ at fixed t' Hooft coupling ($A$ fixed)  for logarithm of the colored HOMFLY polynomials $\ln H_R^{\cal K}(A,q)$.
There is an alternative genus expansion \cite{MMS} known as Hurwitz-Fourier
transform  in variable $\mathfrak{h}$:
\be
H_R^{\cal K}(A,q) = \hbox{dim}_R(A,q)\cdot \Big(\sigma_{_\Box}^{\cal K}(A)\Big)^{|R|}
\cdot \exp \left(\sum_{\Delta}
\mathfrak{h}^{|\Delta|+l(\Delta)-2}s_\Delta^{\cal K}(A,\mathfrak{h} ^2)
\varphi_R(\Delta)\right)\ \ \ \ \ \ \ \ \ \mathfrak{h}={\hbar\over \Big(\sigma_{_\Box}^{\cal K}(A)\Big)^{2}}
\label{HOMFLYhur}
\ee
where   $\sigma_{_\Box}^{\cal K}(A)= H^{\cal K}_{\Box}(q=1,A)$ are called special polynomials.
This expansion results in the appearance of Hurwitz-Tau function when substituted in the
Ooguri-Vafa partition function $Z_{OV}^{\cal K}(A,q,\bar p)$ \cite{MMN,AMMNhurtau}.
Here the sum goes over the Young diagrams $\Delta$ with $l(\Delta)$ lines
of lengths $\delta_i$ and the number of boxes $|\Delta|=\sum_{i}^{l(\Delta)} \delta_i$,
while $\varphi_R(\Delta)$ are proportional to the characters of symmetric groups $\psi_R(\Delta)$ at $|R|=|\Delta|$:
$\psi_R(\Delta)= z_{\Delta}d_R\varphi_R(\Delta)$, and continued to $|R|>|\Delta|$ as in \cite[eq.(3)]{MMN}. Here $d_R$
is the dimension of representation $R$ of the symmetric group $S_{|R|}$ divided by $|R|!$
and $z_{\Delta}$ is the standard symmetric factor of the Young diagram (order of the automorphism)
\cite{Fulton}. With this definition, the sum in (\ref{HOMFLYhur}) runs over all $\Delta$ such that $|\Delta|\le |R|$, while in all sums below we use $\psi_R(\Delta)$ which leaves in sums only terms with $|R|=|\Delta|$. An advantage to use the expansion (\ref{HOMFLYhur}) is in a possibility of lifting $\varphi_R(\Delta)$ to a ring of cut-and-join operators $W_\Delta$ \cite{MMN}, while using the basis of $\psi_R(\Delta)$ is better for studying integrality conjectures.

This Hurwitz version of the Fourier transform in the color index $R$, (\ref{HOMFLYhur})
converts the set of colored HOMFLY polynomials
into a collection of {\it generalized special polynomials}
$\sigma_{g|\Delta}^{\cal K}(A)$ \cite{MMS}.
They enter (\ref{HOMFLYhur}) through
\be
s_\Delta^{\cal K}(A,\mathfrak{h}^2) = \sum_{g\geq 0} \mathfrak{h}^{2g} \sigma_{g|\Delta}^{\cal K}(A)
\ee
Note that the free energy behaves as $\hbar^{-2}$, which is natural for
$\tau$-functions.

The properties of this genus/Hurwitz expansion of individual knot polynomials
did not yet gain enough attention.
However, eq.(\ref{HOMFLYhur}) calls for study of the genus expansion of the Ooguri-Vafa
partition functions and implies that a natural form for it should involve
the plethystic exponential:
\be
Z_{OV}^{\cal K}\{A,q\,|\,\bar p\} = \sum H_R^{\cal K}(A,q)\chi_R\{\bar p\} =
\exp \left(\sum_{d\geq 1} \sum_\Delta  \frac{\prod_{i=1}^{l(\Delta)}(q^{d\delta_i}-q^{-d\delta_i})}{d}
\,S_\Delta^{\cal K}\ (A^d,q^d)  \widehat{Ad}_d (\bar p_{\Delta})
\right)
\label{OVhur}
\ee
where $\bar p_{\Delta}=\prod_i \bar p_i^{\,\mu_i}$ and
$\ \widehat{Ad}_d (\bar p_{\Delta}) =\prod_i \bar p_{di}^{\,\mu_i}$, where $\mu_i$ is equal to the number of times that the line of length $i$ is met in the Young diagram $\Delta$. In these terms, $z_\Delta=\prod_i i^{\mu_i}\mu_i!$\ .

Relation between (\ref{OVhur}) and (\ref{HOMFLYhur}) is not at all naive, since the sum of logarithm is not equal to a logarithm of the sum, or the sum of genus expansions is not the same as a genus expansion of the sum.
It involves generalizations of the Cauchy formula (\ref{ZOVunknot}) to the generation function of the generalized Hurwitz numbers \cite{Dij,Ok,MMN}
\be
Z_{OV}^{\cal K}\{A,q\,|\,\bar p\}\equiv Z_{Hurw}=\sum_R {\rm dim}_{_R}(A,q) \cdot  \chi_R\{\bar p\}\cdot e^{\sum_{\Delta}\beta_{\Delta}\varphi_R(\Delta)}
\ee
where $\beta_{\Delta}$ encodes knot ${\cal K}$ information. The  $\varphi_R(\Delta)$ appears in the ``multipoint "correlators" of generalized symmetric group characters as follows:
\be
{\rm Hurw}_q(\Delta_1,\Delta_2,\ldots,\Delta_m) = \sum_{|R|=q} d_R^2
\varphi_R(\Delta_1)\varphi_R(\Delta_2)\ldots \varphi_R(\Delta_m)
\ee
Note  that (\ref{OVhur}) uses yet another different version of genus expansion, that is.,
in power of $q-q^{-1}$ rather than $\hbar$ \cite{MMS}. One can rewrite the product
\be
{\prod_{i=1}^{l(\Delta)}(q^{d\delta_i}-q^{-d\delta_i})}={\prod_{j=1}(q^{jd}-q^{-jd})^{\mu_j}}
\ee
which resembles the measure for the $\beta$-ensemble \cite{HS5}:
$(x_1-x_2)^\beta \longrightarrow  \prod_{i=0}^{\beta-1}(x_1-q^{2i} x_2)$. One can also look at it as  a product of $q$-numbers $[\delta_i]_{q^d}$ or $[j]_{q^d}^{\mu_j}$, which generates an additional factor of $(q^d-q^{-d})^{l(\Delta)}$.
Hence, the additional suppression $\hbar^{|\Delta|}$
in (\ref{HOMFLYhur}) disappears from (\ref{OVhur}).

The LMOV integrality conjecture claims that after one more Hurwitz transform
of $S_\Delta(q,A)$ in (\ref{OVhur}),
\be\label{gN}
S_\Delta(q,A) = \sum_Q \psi_Q(\Delta)\cdot G_Q(A,q)
\ee
the genus expansions
\be
G_Q(A,q)
= \sum_{g\geq 0, k}N_{Q,g,k} A^k(q-q^{-1})^{2g-2}
\ee
have integer coefficients. Moreover, integers $N_{Q,g,k}$ at fixed $Q$ and $k$ actually vanish
at high enough genus $g$, this is an advantage of the above mentioned invariant version
of the expansion.
This LMOV integrality of every term of the genus expansion
is, of course, much stronger than just integrality of the entire free energy.

Note that relation (\ref{gN}) can be immediately inverted:
\be
G_Q(A,q)=\sum_{\Delta} {1\over z_\Delta}\psi_Q(\Delta)S_\Delta(q,A)
\ee
due to the orthogonality conditions
\be\label{orth}
\sum_R {1\over z_\Delta}\psi_R(\Delta)\psi_{R}(\Delta ')=\delta_{\Delta\Delta '},\ \ \ \ \ \sum_\Delta {1\over z_\Delta}\psi_R(\Delta)\psi_{R'}(\Delta)=\delta_{RR'}
\ee
An important implication of the Hurwitz approach is that $Z_{OV}\{\bar p\}$
should satisfy the AMM/OE topological recursion in $g$,
and this fact was actually used in the study of proofs of LMOV relation in \cite{LP}.
Such studies have not be extended to prove Marino's integrality conjectures
involving Kauffman polynomials which we hope to pursue in future.

It is appropriate to mention that Marino's conjectures have been verified for
some torus knots and links \cite{Marino,Stevan,PBR} and figure-eight knot \cite{NRZ}.
Hence the main focus in this paper is to verify Marino's integrality conjectures for various arborescent knots up to 8 crossings using colored Kauffman polynomials
($SO(N)$ colors up to two boxes in Young diagram) and HOMFLY polynomials for mixed $SU(N)$ representations. The latter are calculated using the universal Racah matrices \cite{MMuniv} which are known only for the arborescent knots.

Informally, arborescent knots are the ones which look like trees with two-bridge branches, i.e. by gluing together the fragments of 4-strand antiparallel braids, see \cite{MMMRV}, where they were named {\it double-fat}. This is what makes their knot polynomials expressible through the two simplest Racah matrices ${\cal S}$ and $\bar {\cal S}$ (see s.3.5 below). They are also well familiar in formal knot theory, see \cite{Con,Cau} for an abstract definition and other details.
The list of arborescent knots includes, in particular, all twisted, 2-bridge and pretzel knots. This means that all prime knots with up to 7 intersections are arborescent being 2-bridge. Among the knots with 8 crossings the only non-arborescent prime knot is $8_{18}$. Moving further in the Rolfsen table, among the knots with 9 crossings, only $9_{34}$, $9_{39}$, $9_{40}$, $9_{41}$, $9_{47}$, $9_{49}$ are non-arborescent. At last, among the knots with 10 crossings, $10_{100}$--$10_{123}$ and $10_{155}$--$10_{165}$ are non-arborescent (polyhedral).

We will also present the LMOV integrality structures for $SU(N)$ colors up to
four boxes in the Young diagram as well. It allows us to reveal a striking feature of the LMOV numbers: it turns out that, with a very high accuracy, the LMOV numbers appeared to be distributed by a Gaussian law as functions of the genus $g$! It is just an observation that has been tested for various knots\footnote{We checked it for all the arborescent knots that have 3-strand braid representation: $3_1$, $4_1$, $5_1$, $5_2$, $6_2$, $6_3$, $7_1$, $7_3$, $7_5$, $8_2$, $8_5$, $8_7$, $8_9$, $8_{10}$, $8_{16}$, $8_{17}$, $8_{18}$, $8_{19}$, $8_{20}$, $8_{21}$ (see s.3.9), for the $T[2,2k+1]$ and $T[3,3k+1]$ series of torus knots, and for the mutant pair $11n41$ and $11n47$.}, and no exception has been found so far.

In sec.\ref{LMOVconj}, we review the  exact formulation of the integrality conjectures.
In sec.\ref{knotpolsurvey}, we briefly recapitulate the recent progress in knot polynomial calculus.
In sec.\ref{intests}, we will present in detail the integrality checks for a particular knot and link.
We refer the reader to our dedicated site \cite{knotebook} where the results for other knots
are updated. In sec.\ref{gauss}, we report on the first application of the obtained results: the Gaussian distribution of the LMOV invariants.
In the concluding sec.\ref{conc}, we summarize the results  obtained.

\section{LMOV conjecture: integrality conditions
\label{LMOVconj}}

\subsection{Integrality conjecture in the HOMFLY case}

As we explained in the introduction section, the genus expansion in knot theory is determined using  gauge/string duality \cite{GV,OV}. That is, $U(N)$ Chern-Simons theory on a three-manifold
$S^3$  is equivalent to topological string theory on a Calabi-Yau manifold which is the resolution of the conifold. The expectation value of the Ooguri-Vafa scalar operator
associated with knots  in $S^3$,
\be\label{OV}
{\cal Z}_{SU(N)}^{\cal K}\{A,q\,|\,\bar p\} = \sum_R H_R^{\cal K}(A,q) \cdot\chi_R\{\bar p\}
\ee
which is a generating function of the unreduced HOMFLY polynomials $H_R^{\cal K}(A,q)$, results in $A$-model  open topological string partition function.
In fact the logarithm of the operator  can be interpreted as
``connected'' correlators $f_R(q,A)$ as follows:
\be\label{Zf}
\log {\cal Z}^{\cal K}_{SU}\{A,q|\bar p\}
= \sum_R \sum_{d=1}^\infty \,\frac{1}{d}{  f}_R^{\cal K}(A^d,q^d) \cdot \widehat{Ad}_d\, \chi_R\{\bar p\}
\ee
We call the new quantities $f_R$  {\it plethystic transforms of the adjoint HOMFLY polynomials} by the reasons explained in the introduction section. Note that, the relation (\ref{gN}) can be inverted leading to  constructing the inverse of plethysm transformation as performed in \cite{LM2}:
\be
f_R^{\cal K}(A,q)=\sum_{d,m=1}(-1)^{m-1}{\mu (d)\over md}\sum_{\Delta_1,\ldots\Delta_m} \widehat{Ad}_d\,\psi_R\Big(\sum_{i=1}^m\Delta_i\Big)\cdot\sum_{R_1,\ldots,R_m} \prod_{j=1}^m{\psi_{R_j}(\Delta_j)\over z_{\Delta_j}}H_{R_j}^{\cal K}(A^d,q^d)
\ee
where the sum of two Young diagrams $\Delta$ and $\Delta'$ is the Young diagram with the lines $\{\delta_i,\delta_i'\}$ with a proper reordering, $\widehat{Ad}_d\,\Delta=\widehat{Ad}_d\,\{\delta_i\}=\{d\delta_i\}$, and $\mu(d)$ is the M\"obius function defined as follows: if the prime decomposition of $d$ consists of $m$ multipliers and contains non-unit multiplicities, $\mu(d)=0$, otherwise $\mu(d)=(-1)^m$.
For representations up to four boxes in Young diagram, the explicit form of the above equation will be
{\footnotesize
\be
f_{[1]}&=&H_{[1]}(q,A)\nn \\
f_{[2]}&=&H_{[2]}(q,A)-{1\over 2}\Big( H_{[1]}(q,A)^2+H_{[1]}(q^2,A^2)\Big)\nn\\
f_{[1^2]}&=&H_{[1^2]}(q,A)-{1\over 2}\Big( H_{[1]}(q,A)^2-H_{[1]}(q^2,A^2)\Big)\nn\\
f_{[3]}&=&H_{[3]}-H_{[2]}H_{[1]}+\frac{1}{3}H_{[1]}^3-\frac{1}{3}H_{[1]}(A^3,q^3)\nn\\
f_{[2,1]}&=&H_{[2,1]}-H_{[2]}H_{[1]}-H_{[1^2]}H_{[1]}+\frac{2}{3}H_{[1]}^3-\frac{1}{3}H_{[1]}(A^3,q^3)\nn\\
f_{[1^3]}&=&H_{[1^3]}-H_{[1^2]}H_{[1]}+\frac{1}{3}H_{[1]}^3-\frac{1}{3}H_{[1]}(A^3,q^3)\nn\\
f_{[4]}&=&H_{[4]}-H_{[3]}H_{[1]}+H_{[2]}H_{[1]}^2-\frac{1}{2}H_{[2]}^2-\frac{1}{4}H_{[1]}^4- \frac{1}{2}H_{[2]}(A^2,q^2)+\frac{1}{4}H_{[1]}^2(A^2,q^2)\nn\\
f_{[3,1]}&=&H_{[3,1]}-H_{[3]}H_{[1]}-H_{[2,1]}H_{[1]}-H_{[2]}H_{[1^2]}+2H_{[2]}H_{[1]}^2+H_{[1^2]}H_{[1]}^2-\frac{1}{2}H_{[2]}^2- \frac{3}{4}H_{[1]}^4+\frac{1}{2}H_{[2]}(A^2,q^2)-\frac{1}{4}H_{[1]}^2(A^2,q^2)\nn\\
f_{[2^2]}&=&H_{[2^2]}-H_{[2,1]}H_{[1]}+H_{[2]}H_{[1]}^2+H_{[1^2]}H_{[1]}^2-\frac{1}{2}H_{[2]}^2-\frac{1}{2}H_{[1^2]}^2- \frac{1}{2}H_{[1]}^4-\frac{1}{2}H_{[2]}(A^2,q^2)-\frac{1}{2}H_{[1^2]}(A^2,q^2)+\frac{1}{2}H_{[1]}^2(A^2,q^2)\nn\\
f_{[2,1^2]}&=&H_{[2,1^2]}-H_{[1^3]}H_{[1]}-H_{[2,1]}H_{[1]}-H_{[2]}H_{[1^2]}+2H_{[1^2]}H_{[1]}^2+H_{[2]}H_{[1]}^2- \frac{1}{2}H_{[2]}^2-\frac{3}{4}H_{[1]}^4+\frac{1}{2}H_{[1^2]}(A^2,q^2)-\frac{1}{4}H_{[1]}^2(A^2,q^2)\nn\\
f_{[1^4]}&=&H_{[1^4]}-H_{[1^3]}H_{[1]}+H_{[1^2]}H_{[1]}^2-\frac{1}{2}H_{[1^2]}^2-\frac{1}{4}H_{[1]}^4- \frac{1}{2}H_{[2]}(A^2,q^2)+\frac{1}{4}H_{[1]}^2(A^2,q^2)\nn
\ee}
Now, the expansion of HOMFLY polynomial is translated into a similar expansion of the plethystic polynomials:
\be
f_R(q,A)=\sum_{n,k} \widetilde {\bf N}_{R,n,k}{A^nq^k\over q-q^{-1}}
\ee
Even though HOMFLY behaves as  $1/(q-q^{-1})^{|R|}$, the reformulated invariant $f_R(q,A)$ has only singularity $1/(q-q^{-1})$

In fact, the only way to check this duality between Chern-Simons and topological string theories is to establish the integrality condition: the coefficients $\widetilde {\bf N}_{R,n,k}$  has to be integer in accordance with the Ooguri-Vafa conjecture \cite{OV}. Moreover, as we explained in the introduction section, one can construct even more refined integers (\ref{gN}) \cite{LMV}
\be\label{ic}
f_R(q,A)=\sum_{n,k\ge 0,Q} C_{RQ}{\bf N}_{Q,n,k}A^n(q-q^{-1})^{2k-1}
\ee
where
\be
C_{RQ}=\sum_{\Delta}{1\over z_\Delta}\psi_R(\Delta)\psi_Q(\Delta){\prod_{i=1}^{l(\Delta)}\Big(q^{\delta_i}-q^{-\delta_i}\Big)\over q-q^{-1}}={1\over q-q^{-1}}\sum_{\Delta}{1\over z_\Delta}\psi_R(\Delta)\psi_Q(\Delta)\ ^*p_\Delta
\ee
with $^*p_k\equiv q^k-q^{-k}$. To compare this formula with (\ref{gN}), one has to use the identity
\be
\sum_R \psi_R(\Delta)\chi_R(p)=p_{\Delta}\nn
\ee
This matrix can be easily reversed using (\ref{orth}):
\be
\Big(C^{-1}\Big)_{QR}=\sum_\Delta {1\over z_\Delta}\psi_R(\Delta)\psi_Q(\Delta){q-q^{-1}\over ^*p_\Delta}
\ee
Let us note that there is a hierarchy of integralities: the weakest statement is the claim that
the HOMFLY polynomials are integer.
The next level is integrality of $\widetilde {\bf N}_{R,n,k}$, which implies that of HOMFLY, but not vice versa.
However, $\widetilde {\bf N}_{R,n,k}$ are not independent: they satisfy some relations \cite{LM1},
while the more refined ${\bf N}_{R,n,k}$ are independent numbers and their integrality
implies the integrality of $\tilde {\bf N}_{R,n,k}$, but not vice versa.
This means that ${\bf N}_{R,n,k}$ are, in a sense, elementary building blocks.
Their integrality from the knot theory point of view is
not at all evident.
Note that the BPS invariants ${\bf N}_{R,n,k}$ are linearly related to the Gopakumar-Vafa
(open Gromov-Witten) invariants ${\bf n}_{\Delta,n,k}$ \cite{GV}:
\be
{\bf n}_{\Delta,n,k}=\sum_{R}\psi_R(\Delta){\bf N}_{R,n,k},\ \ \ \ \ \ \ \ {\bf N}_{R,n,k}=\sum_{\Delta} {1\over z_\Delta}\psi_R(\Delta){\bf n}_{\Delta,n,k}
\ee
and the integrality of ${\bf n}_{\Delta,n,k}$ follows from the integrality of ${\bf N}_{R,n,k}$, but not vice versa.
Integrality of the coefficients ${\bf N}_{Q,n,k}$ was checked in \cite{RS,BRS,RZ}, and was also generally proven in \cite{LP}. However, as an illustration, we have calculated all ${\bf N}_{Q,n,k}$ with $|Q|\le 4$ for the knots in the Rolfsen table \cite{katlas} given by 3-strand braids and manifestly checked their integrality. We  discuss this in sec.\ref{Nf}.

\paragraph{Framing dependence.}
Explicit answers for the functions $f_R(q,A)$ and, hence, for all the integers
depend on the choice of framing.
Remarkably, this dependence can be described by the action of the cut-and-join operator
and {\it almost does not} affect the integrality:
the integers remain integers \cite{MV} at any framing with a small additional rescaling of the HOMFLY polynomials entering the definition (\ref{OV}),
though the dependence of integers on the framing is quite weird (see examples in \cite[sec.4.3]{MV}\footnote{Notice a misprint in Table 8 of \cite[sec.4.3]{MV}: in the second line of the table, there should be $-8-5p-3p^2$ instead of $-8-5p-p^2$.}).
The total framing factor contains two multipliers: $A^{p|R|}$ and $q^{2p\varphi_R([2])}$
($\varphi_R([2])$ proportional to the quadratic Casimir), where $p$ is an arbitrary integer. The first multiplier is trivial, since it is removed by the replace $\bar p_k\to A^p\bar p_k$ in (\ref{Zf}), i.e. leads to a trivial factor of $A^{p|R|}$ in $f_R(q,A)$. The second factor is much less trivial. What is more important, in order to preserve the integrality, one has to change the definition (\ref{OV}) making it slightly dependent on framing: one should multiply the HOMFLY polynomials  entering it by an additional framing factor: $H_R^{\cal K}(A,q)\to (-1)^{p|R|}H_R^{\cal K}(A,q)$, \cite{AV}.

In fact, the framing story is different for knots and links.
For knots, there is a distinguished {\it topological framing} (standard framing),
and we present all the answers below for {\it this} choice.
For links, there is no distinguished "mutual" framing of different components of the link.
Moreover, here one should additionally care that the HOMFLY and Kauffman polynomials are calculated
in the same framing. Another subtlety is a factor that distinguish between the reduced and unreduced knot polynomials. While in the HOMFLY case they differ just by the corresponding quantum dimensions, the standard Kauffman polynomial of a link is related with the unreduced one by multiplying with the quantum dimension {\it and} with a factor of $A^{-2 {\rm lk} ({\cal L})}$, where ${\rm lk} ({\cal L})$ is the linking number.

Note that in order to fix notation in the case of links, one can use another distinguished framing, the
{\it vertical framing}, which means that all ${\cal R}$-matrices are generated from the
{\it universal} one.
This prescription fixes the notation, but it is {\it different} from the topological framing for knots.
Fortunately, the relation in this case is very simple: for knots
$H^{\cal K,{\rm top}}_R = A^{-w|R|} q^{-4w\varphi_R([2])} \cdot H^{\cal K,{\rm vert}}_R$,
where $w$ is the writhe number.

\subsection{Integrality conjectures in the Kauffman case}
A natural generalization of the  described correspondence is the equivalence between the $SO/Sp$ Chern-Simons theory and the topological string theory  on an orientifold of the small resolution of the conifold \cite{SV}-\cite{NRZ}. In this case, the Chern-Simons partition function is associated with two types of contributions: those from oriented and non-oriented strings, the former ones coming with the degree 1/2:\footnote{This formula looks quite natural due to the Rudolph-Morton-Ryder theorem, \cite{RMR}:
\be
\Big(K^{\cal K}_R\Big)^2=H_{R,R}\ \ \ \ \ \ \hbox{mod }2\nn
\ee
where "mod 2" means that the integer coefficients of the Laurent polynomials in this formula are taken modulo 2, $K_R$ is the Kauffman knot polynomial and $H_{R,S}$ denotes the HOMFLY polynomial in the composite representation \cite{Koike}. In fact, the Rudolph-Morton-Ryder theorem immediately follows from the integrality conditions, see \cite{Marino}.
}
\be
{\cal Z}^{\cal K}_{SO/Sp}\{A,q|\bar p\}=Z_{no}\sqrt{Z_o}
\ee
Thus, one expects that
\begin{itemize}
\item The partition function of the oriented strings $Z_{o}$ induces an integrality condition.
\item The partition function of the non-oriented strings $Z_{no}$ also induces another integrality condition.
\end{itemize}
We will now briefly review the necessary steps:
The oriented partition function is given by the generating function of the HOMFLY polynomials in
composite representations \cite{Marino}:
\be
Z_o\{A,q\,|\,\bar p\} =
\sum_{R,S} H_{(R,S)}^{\cal K}(A,q)\cdot\chi_R\{\bar p\}\chi_S\{\bar p\}
\label{ZHRS}
\ee
Note that the sum in the second case is over a double set of Young diagrams,
but there is a single set of time variables $\bar p$.
From this partition function one builds the free energy, which is again
expanded into sum over Young diagrams,
\be\label{Zo}
\log Z_o\{A,q|\bar p\}
= \sum_R \sum_{d=1}^\infty \,\frac{1}{d}{  h}_R^{\cal K}(A^d,q^d) \cdot \widehat{Ad}_d\, \chi_R\{\bar p\}
\ee
and the first few terms of expansion are
\be
{  h}^{\cal K}_{[1]}(A,q) &=& 2H_{[1]}(A,q) \nn \\
{  h}^{\cal K}_{[2]}(A,q) &=& 2H_{[2]}(A,q)+H_{([1],[1])}(A,q)
- 2\Big(H_{[1]}(A,q)\Big)^2 - H_{[1]}(A^2,q^2) \nn \\
{  h}^{\cal K}_{[1,1]}(A,q) &=& 2H_{[1,1]}(A,q)+H_{([1],[1])}(A,q)
- 2\Big(H_{[1]}(A,q)\Big)^2 + H_{[1]}(A^2,q^2) \nn \\
\ldots
\ee
There is a mirror symmetry under transposition of Young diagrams which  relates knot polynomials as follows:
\be
H_{R^{tr}}(A,q) = H_R(A,-q^{-1})
\label{traninvers}
\ee
However its implication to $h_R$ is not seen.
Note that the sign flip in (\ref{traninvers}) emerges in the course of performing the Adams transformation in (\ref{Zo}).

One can again generate the refined integrality condition via
\be\label{icK1}
h_R(q,A)=\sum_{n,k\ge 0,Q} C_{RQ}\hat {\bf N}^{c=0}_{Q,n,k}A^n(q-q^{-1})^{2k-1}
\ee
where the superscript $c$ denotes the contribution from Riemann surfaces with $c$ cross-cups \cite{BFM,BR}.

In order to calculate the non-oriented partition function, one has to calculate
\be
Z_{no}={{\cal Z}^{\cal K}_{SO/Sp}\{A,q|\bar p\}\over\sqrt{Z_o}}
\ee
where the numerator is given by the generating function of the (unreduced) Kauffman polynomials
\be
{\cal Z}_{SO/Sp}^{\cal K}\{A,q|\bar p\} = \sum_R K_R^{\cal K}(A,q) \cdot\chi_R\{\bar p\}
\ee
and $Z_o$ is given by the HOMFLY polynomials in composite representations, (\ref{ZHRS}). Hence,
\be
\log Z_{no}\{A,q|\bar p\}=\log {\cal Z}^{\cal K}_{SO/Sp}\{A,q|\bar p\}-{1\over 2}\log Z_o\{A,q|\bar p\}
= \sum_R \sum_{d\ge 1,\ odd}^\infty \,\frac{1}{d}{  g}_R^{\cal K}(A^d,q^d) \cdot \widehat{Ad}_d\, \chi_R\{\bar p\}
\ee
with the first terms of expansion being
\be
{  g}^{\cal K}_{[1]}(A,q)& =& K_{[1]}(A,q)-H_{[1]}(A,q) \nn \\
{  g}^{\cal K}_{[2]}(A,q)& =& K_{[2]}(A,q)- {1\over 2}\Big(K_{[1]}(A,q)\Big)^2-H_{[2]}(A,q)
+\Big(H_{[1]}(A,q)\Big)^2-{1\over 2}H_{([1],[1])}(A,q) \nn \\
{  g}^{\cal K}_{[1,1]}(A,q) &=& K_{[1,1]}(A,q)- {1\over 2}\Big(K_{[1]}(A,q)\Big)^2-H_{[1,1]}(A,q)+\Big(H_{[1]}(A,q)\Big)^2-{1\over 2}H_{([1],[1])}(A,q)  \nn \\
\ldots\nn
\ee
and the integrality condition
\be\label{icK2}
g_R(q,A)=\sum_{n,k\ge 0,Q} C_{RQ}\Big(\hat {\bf N}^{c=1}_{Q,n,k}A^n(q-q^{-1})^{2k}+\hat {\bf N}^{c=2}_{Q,n,k}A^n(q-q^{-1})^{2k+1}\Big)
\ee
\bigskip
The above discussion for knots can be extended to two component links. The relevant operator for these links will be
\be
Z_o\{A,q\,|\,p,\bar p\} &=&
\sum_{R,S} H_{(R_1,S_1)(R_2,S_2)}^{\cal L}(A,q)\cdot\chi_{R_1}\{p\}\chi_{S_1}\{p\}\cdot\chi_{R_2}\{\bar p\}\chi_{S_2}\{\bar p\}
\\
\log Z_o\{A,q|p,\bar p\}
&=& \sum_R \sum_{d=1}^\infty \,\frac{1}{d}{  h}_{R_1,R_2}^{\cal L}(A^d,q^d) \cdot \widehat{Ad}_d\, \chi_{R_1}\{p\}\cdot \widehat{Ad}_d\, \chi_{R_2}\{\bar p\}
\ee
and
\be
{\cal Z}_{SO/Sp}^{\cal L}\{A,q|p,\bar p\}&=&
\sum_{R_1,R_2} K_{R_1,R_2}^{\cal L}(A,q)\cdot\chi_{R_1}\{p\}\chi_{R_2}\{\bar p\}\nn
\\
\log Z_{no}\{A,q|p,\bar p\}&=&\log {\cal Z}^{\cal L}_{SO/Sp}\{A,q|p,\bar p\}-{1\over 2}\log Z_o\{A,q|p,\bar p\}\nn\\
&=& \sum_R \sum_{d\ge 1,\ odd}^\infty \,\frac{1}{d}{  g}_{R_1,R_2}^{\cal K}(A^d,q^d) \cdot \widehat{Ad}_d\, \chi_{R_1}\{p\}\cdot \widehat{Ad}_d\, \chi_{R_2}\{\bar p\}
\ee
so that the explicit form for oriented invariants $h_{R_1,R_2}$ for some representations are
\be\label{hL}
{h}^{\cal L}_{[1],[1]}&=&2H^{\mathcal{L}}_{[1],[1]}+2H^{\bar{\cal L}}_{[1],[1]}-4H^{{\cal K}_1}_{[1]}H^{{\cal K}_2}_{[1]}\\
{h}^{\cal L}_{[2],[1]}&=&2H^{{\cal L}}_{[2],[1]}+2H^{\bar{\cal L}}_{[2],[1]}+2H^{{\cal L}}_{([1],[1]),[1]}-4H^{{\cal L}}_{[1],[1]}H^{{\cal K}_1}_{[1]}-4H^{\bar{\cal L}}_{[1],[1]}H^{{\cal K}_1}_{[1]}\nn\\
&&-4H^{{\cal K}_1}_{[2]}H^{{\cal K}_2}_{[1]}-2H^{{\cal K}_1}_{([1],[1])}H^{{\cal K}_2}_{[1]}+8\left(H^{{\cal K}_1}_{[1]}\right)^2H^{{\cal K}_2}_{[1]}\nn\\
{h}^{\cal L}_{[1,1],[1]}&=&2H^{{\cal L}}_{([1,1],[1]}+2H^{\bar{\cal L}}_{[1,1],[1]}+2H^{{\cal L}}_{([1],[1]),[1]}-4H^{{\cal L}}_{[1],[1]}H^{{\cal K}_1}_{[1]}-4H^{\bar{\cal L}}_{[1],[1]}H^{{\cal K}_1}_{[1]}\nn\\
&&-4H^{{\cal K}_1}_{[1,1]}H^{{\cal K}_2}_{[1]}-2H^{{\cal K}_1}_{([1],[1])}H^{{\cal K}_2}_{[1]}+8\left(H^{{\cal K}_1}_{[1]}\right)^2H^{{\cal K}_2}_{[1]}\nn\\
\ldots\nn
\ee
and similarly
\be\label{gL}
{  g}^{\cal L}_{[1],[1]}&=&K_{[1],[1]}^{\cal L}-K_{[1]}^{{\cal K}_1}K_{[1]}^{{\cal K}_2}- H_{[1],[1]}^{\cal L}-H_{[1],[1]}^{\bar{\cal L}}+2H_{[1]}^{{\cal K}_1}H_{[1]}^{{\cal K}_2}\\
g_{[2],[1]}^{\cal L}&=&K_{[2],[1]}^{\cal L}-H_{[2],[1]}-H_{[2],[1]}^{\cal L}-H_{[2],[1]}^{\bar{\cal L}}- K_{[1],[1]}^{\cal L}K_{[1]}^{{\cal K}_1}- K_{[2]}^{{\cal K}_1}K_{[1]}^{{\cal K}_2} +2H_{[1],[1]}^{\cal L} H_{[1]}^{{\cal K}_1}\nn\\
&~&+2H_{[1],[1]}^{\bar {\cal L}}H_{[1]}^{{\cal K}_1}+
2H_{[2]}^{{\cal K}_1}H_{[1]}^{{\cal K}_2} +H_{([1],[1])}^{{\cal K}_1}H_{[1]}^{{\cal K}_2} +\Big(K_{[1]}^{{\cal K}_1}\Big)^2K_{[1]}^{{\cal K}_2} -4\Big(H_{[1]}^{{\cal K}_1}\Big)^2H_{[1]}^{{\cal K}_2}\nn\\
g_{[1,1],[1]}^{\cal L}&=&K_{[1,1],[1]}^{\cal L}-H_{[1,1],[1]}-H_{[1,1],[1]}^{\cal L}-H_{[1,1],[1]}^{\bar{\cal L}}- K_{[1],[1]}^{\cal L}K_{[1]}^{{\cal K}_1}- K_{[1,1]}^{{\cal K}_1}K_{[1]}^{{\cal K}_2} +2H_{[1],[1]}^{\cal L} H_{[1]}^{{\cal K}_1}\nn\\
&~&+2H_{[1],[1]}^{\bar {\cal L}}H_{[1]}^{{\cal K}_1}+
2H_{[1,1]}^{{\cal K}_1}H_{[1]}^{{\cal K}_2} +H_{([1],[1])}^{{\cal K}_1}H_{[1]}^{{\cal K}_2} +\Big(K_{[1]}^{{\cal K}_1}\Big)^2K_{[1]}^{{\cal K}_2} -4\Big(H_{[1]}^{{\cal K}_1}\Big)^2H_{[1]}^{{\cal K}_2}\nn\\
\ldots \nn
\ee
where ${\cal K}_1$ and ${\cal K}_2$ are the components of the link and $\bar{\cal L}$ denotes the link obtained from ${\cal L}$ by reversing the orientation of one of its components.
These expansions are again related to the two corresponding integrality conditions:
\be\label{icKl1}
h_{R_1,R_2}(q,A)=\sum_{n,k\ge 0,Q_1,Q_2} C_{R_1Q_1}C_{R_2Q_2}\hat {\bf N}^{c=0}_{Q_1,Q_2,n,k}A^n(q-q^{-1})^{2k}
\ee
and
\be\label{icKl2}
g_{R_1,R_2}(q,A)=\sum_{n,k\ge 0,Q_1,Q_2} C_{R_1Q_1}C_{R_2Q_2}\Big(\hat {\bf N}^{c=1}_{Q_1,Q_2,n,k}A^n(q-q^{-1})^{2k+1}+\hat {\bf N}^{c=2}_{Q_1,Q_2,n,k}A^n(q-q^{-1})^{2k+2}\Big)
\ee
Note that, in the case of link, $h$ and $g$ have to contain additional degrees of $q-q^{-1}$ at $q\to 1$ as compared with the knot case.
The integrality expansions reviewed for unoriented topological string amplitudes were conjectured \cite{Marino} and verified for
$(2,2m+1)$ torus knots. Now with our recent advances in evaluation of colored knot polynomials for adjoint representations
for arborescent knots and non-arborescent knots obtained from three strand braids,
\cite{MMMRV}-\cite{MMMSsu}, \cite{MMuniv},  we could provide further evidence for the conjecture.

The main goal of the present paper is to determine the coefficients $N$ for a wide class of knots/links and check their integrality properties. Our results for many knots and links can be considered as  a direct continuation of the Appendix from \cite{IMMMIII}
and especially of Appendix B from \cite{NRZ} for figure-eight knot
where the theory  is presented in detail with relevant references.

In the following section we will present briefly various methods useful in the evaluation of colored polynomials.

\section{Colored polynomials for arborescent knots
\label{knotpolsurvey}}
The colored HOMFLY polynomials are well defined quantities.
If a link/knot is presented as the closure of a braid,
then the HOMFLY polynomial is a $q^\rho$-weighted trace of a product of quantum ${\cal R}$-matrices
at the intersections of strands of the braid \cite{RT}.
In the modern version of the RT formalism \cite{MMMkn12}-\cite{RTmod2}, one uses the ${\cal R}$-matrices
acting in the space of intertwining operators.
Actually this defines the HOMFLY polynomial up to an overall framing factor.
For knots there is a distinguished choice of framing called  {\it the topological framing}
which is independent of framing number.
Remarkably, such a framing choice is not necessary for LMOV integrality structure. These properties hold for other framings
where we add a suitable $U(1)$ invariant with suitable $U(1)$ charges \cite{MV}.
However, for links the distinguished framing does not exist.  Moreover, there is an additional
ambiguity in HOMFLY depending on mutual orientation of components.

Despite these constraints, the colored HOMFLY polynomials are very difficult to {\it evaluate},
and we have very limited success  in this direction for arborescent knots and links.
The main barrier in obtaining polynomial form  is the absence of $SU(N)$ Racah matrices in quantum group theory.
Finding these Racah matrices for arbitrary representation gets especially difficult in the case of non-trivial multiplicities
(i.e. for non-rectangular Young diagrams and outside the $E_8$-sector \cite{MMkrM}). Such representations
with non-trivial multiplicity plays a crucial role in distinguishing {\it mutant} knot pairs \cite{MMMRV}-\cite{MMMSsu}.
In the following subsection, we will briefly review various methods leading to knot polynomials.

\subsection{Inclusive Racah matrices for 3-strand braids}
The brute force application of the modern RT formalism {\it a la} \cite{MMMkn12}-\cite{RTmod2} requires knowledge of
the matrices ${\cal R}_{a,a+1}$ acting at the crossing of adjacent strands $a$ and $a+1$ in the braid.
While one of them, say ${\cal R}_{12}$, can be diagonalized and has very simple eigenvalues,
which are just exponentials of quadratic Casimir eigenvalues $\varkappa_Y$ \cite{DMMSS},
the others are not diagonal and are obtained by conjugation with additional {\it mixing} matrices.
In particular, ${\cal R}_{23} = {\cal U}{\cal R}_{12} {\cal U}^\dagger$,
where ${\cal U}$ is a Racah matrix converting an intertwiner
$(R\otimes R)\otimes R \longrightarrow Q\in R^{\otimes 3}$
into $R\otimes (R \otimes R) \longrightarrow Q$.
It is a matrix acting in the space of representations $Y\in R^{\otimes 2}$.
Thus the knowledge of the {\it inclusive} Racah matrix, i.e. a collection of Racah matrices
for all $Q\in R^{\otimes 3}$ is sufficient for performing the 3-strand braid calculations.
Going beyond three strand braid required determining a wider class of inclusive Racah matrices which is tedious.

\subsection{Highest weight method}
This method gives a straightforward evaluation of mixing matrices which requires comparison
of linear bases inherited from the decompositions $R^{\otimes(\#\ {\rm of\ strands})}$
with different order of brackets, like
$R^{\otimes 2}\otimes R$ and $R\otimes R^{\otimes 2}$.
Literally, if the vector spaces are associated with particular representations,
this comparison gives the Clebsh-Gordon coefficients.
In order to get the Racah matrices, the simplest way is to look just at the
highest weight vectors as elements in the abstract Verma modules.
This formalism is successfully developed in \cite{MMMS21} and \cite{MMMS31}
and has already allowed us to find the {\it inclusive} Racah matrices for $R=[2,2]$ and even $R=[3,1]$.
In combination with the differential expansion method \cite{DGR,IMMMfe,MMMevo,diffarth,Naw,Konodef,Mor},
this provides extensions to other rectangular representations.
Further progress (for other non-rectangular representations) is expected after developing the $\Delta$-technique
briefly outlined in \cite{MMMS31}.
We are presently extending the work \cite{IMMMIII} investigating the highest weight method to determine  polynomials of knots obtained from four or more strands carrying symmetric representation.

Even though  the method is straightforward  and very successful, the calculations become cumbersome as we increase the number of strands beyond three strands.


\subsection{Eigenvalue hypothesis}

The most interesting method is the eigenvalue hypothesis \cite{IMMMev}
saying that the entries of Racah matrix are actually made from the known eigenvalues
$\pm q^{\varphi_Y([2])}$
of ${\cal R}$-matrix for all representations $Y\in R^{\otimes 2}$
(the sign depends on belonging to the symmetric or anti-symmetric squares).
Explicit formulas are currently known up to the size $6\times 6$ (see \cite{IMMMev}, \cite{MMuniv} and \cite{MMMSsu}),
while  for $R=[3,1]$ Racah matrices can be $20\times 20$.
Still, most of constituents of the inclusive Racah matrices are small, and the use of eigenvalue hypothesis is practically very convenient even in its present form.
However there are conceptual questions \cite{Wenzl} that still need to be resolved within this method.

\subsection{Sum over paths for fundamental representations and cabling}

A natural way is to represent ${\cal R}$-matrices in the space of paths
in the representation tree, which leads to a peculiar sum-over-paths formulation,
at least, for the fundamental HOMFLY \cite{Anopath}.
Then, the cabling method can be applied to extract the colored HOMFLY polynomials \cite{RTmod2}. This method turns out to be rather powerful  and calculations
involving 12-strands determine  $[21]$-colored HOMFLY polynomials for some 4-strand knots,
and  $[31]$- or $[22]$-colored HOMFLY polynomials for the 3-strand braids.

\subsection{Two bridge and other arborescent (double-fat) knots}

A big class, the arborescent knots \cite{Con,Cau}, which dominate in the Rolfsen table of knots with low crossing numbers, has a peculiar double-fat realization \cite{MMMRV},
which expresses their HOMFLY polynomials through just two {\it exclusive} Racah matrices
${\cal S}:\ \ \Big\{(R\otimes R)\otimes \bar R \longrightarrow R\Big\}
\ \longrightarrow \ \Big\{ R\otimes (R\otimes \bar R) \longrightarrow R\Big\}\ $
and
$\ \ \bar {\cal S}:\ \ \Big\{(R\otimes \bar R)\otimes  R \longrightarrow R\Big\}
\ \longrightarrow \ \Big\{ R\otimes (\bar R\otimes R) \longrightarrow R\Big\}$.
The term {\it exclusive} refers to selecting just one particular representation $R$
from the product $R^{\otimes 2}\otimes \bar R$.
{\it Exclusive} is, of course, much simpler than {\it inclusive}; however,
involvement of the conjugate representation (inverted strand direction), is a considerable
complication.
The matrices ${\cal S}$ and $\bar{\cal S}$ are known for all symmetric (and antisymmetric) representations $R$
\cite{MMSpret} and \cite{nara} and, by an outstanding effort, for $R=[2,1]$ \cite{GuJ}.

\subsection{${\cal S}$ and $\bar{\cal S}$ from exclusive Racah}

A much simpler way to obtain ${\cal S}$ and $\bar{\cal S}$ for non-symmetric representations
was suggested in \cite{MMMSsu}.
Namely, the exclusive Racah matrices ${\cal S}$ were extracted from the HOMFLY polynomials of the double evolution family \cite{MMMevo}
of 3-strand braids (which were evaluated with the known inclusive Racah matrices) in the following way: the same family can be presented as an arborescent family (of the pretzel knots), hence, its HOMFLY polynomials can be presented in the form involving the exclusive matrices in such a way that ${\cal S}$ diagonalizes the double evolution matrix. Then the second exclusive matrix $\bar{\cal S}$ is obtained from the relation
\be
\bar{\cal S} = \bar T^{-1} {\cal S} T^{-1} {\cal S}^\dagger \bar T^{-1}
\label{SvsbS}
\ee
which is always correct for the Racah matrices \cite{GuJ} and follows from triviality of two unlinked unknots \cite{MMMRV}.

\subsection{Families of arborescent knots}

This approach to the exclusive Racah matrices is yet another impressive success of the evolution method \cite{DMMSS,MMMevo}, which describes each knot or link together with a whole family
which arises when any of the encountered ${\cal R}$-matrices is raised to an arbitrary power.
The point is that calculations for the entire family is technically the same,
but one obtains this way the HOMFLY polynomial for many knots at once, and also a new parameter,
in which interesting recursions, of course, immediately arise.
Most important, this provides a new ordering in the space of knots, which unifies
knots of a similar {\it complexity}, which has nothing to do with the number of crossings
used in the Rolfsen table.
First examples of this {\it family method} application
are provided in \cite{MMfam} and \cite{MMMRSS}.
The actual tabulation of colored knot polynomials in \cite{knotebook}, basing on
\cite{MMMRV}-\cite{MMMSsu} was made possible only by use of this method.

\subsection{Universal knot polynomials}

It is not easy to include conjugate representations, which will involve the rank $N$ dependence, within highest weight method which is $N$-independent. Interestingly for adjoint representations,
Vogel's universality hypothesis \cite{Vog} claims that they can be
formulated in a {\it universal}, group-independent way.
The hypothesis actually originated from knot theory studies, and the idea was to raise it
up to the group theory level, where it partly failed.
However, not very surprisingly, the knot polynomials are not sensitive to the failures,
and they are indeed {\it universal} \cite{MMkrM,MMuniv}.
Moreover, an extension of Vogel's hypothesis from the dimensions and Casimirs to the Racah matrices,
which is one of the steps required for evaluating the adjoint HOMFLY polynomial,
also provided the non-trivial confirmation of the {\it eigenvalue hypothesis} and
explicit formulas for the $6\times 6$ Racah matrices \cite{MMuniv}.
This data  has been useful for writing colored HOMFLY and colored Kauffman  for adjoint representation.

\subsection{A collection of colored knot polynomials}

As we cited throughout this paper, the data on the colored knot polynomials are collected in the website \cite{knotebook}. We briefly describe here the structure of this site.

Basically, it consists of three large parts (apart from links to other knot tables that contain only uncolored knot polynomials with a notable exception of the colored Jones ones): the first part contains a description of some important families of knots and links; the second one contains a theoretical part with links to papers useful for evaluation of colored polynomials; and, the most important third part contains the data: the Racah and mixing matrices, which allow one to evaluate knot polynomials, and the results of this evaluation.

The results are: (i) for the HOMFLY polynomials \cite{kb1}, (ii) for the Kauffman polynomials \cite{kb2} and (iii) for the universal polynomials \cite{kb3}. Most of the results are either for knots from the Rolfsen table or for special families of knots. Since the case of  HOMFLY polynomials is most developed, section (i) contains, apart from the polynomials themselves some additional information about the structure of the answers mostly related with their differential expansion. At last, this section contains the LMOV integers up to the fourth level for all arborescent knots from the Rolfsen table with no more than 8 crossings which have a 3-strand braid representation. These are the knots: $3_1$, $4_1$, $5_1$, $5_2$, $6_2$, $6_3$, $7_1$, $7_3$, $7_5$, $8_2$, $8_5$, $8_7$, $8_9$, $8_{10}$, $8_{16}$, $8_{17}$, $8_{18}$, $8_{19}$, $8_{20}$, $8_{21}$. Similarly, section (ii), which has been less developed yet contains, apart from the Kauffman polynomials only the LMOV integers up to level two for the same set of knots. These LMOV tables are exactly the data obtained as a result of the present paper.

\section{Tests of integrality conjectures
\label{intests}}

\subsection{$SU(N)$ Chern-Simons\label{Nf}}

As discussed in the introduction, the integrality conjecture (\ref{ic}) famously known as LMOV condition has been proven in \cite{LP}. Our focus in this paper is to write  integer coefficients for the representations $Q$ with $|Q|\le 4$. The HOMFLY polynomials for these representations and a list of the integers for more knots from the Rolfsen table \cite{katlas} can be found in \cite{knotebook}. Here we present for a knot $8_{20}$ from the Rolfsen table. This knot $8_{20}$ is an arborescent and can also be obtained from 3-strand braid. The reason for our choice is that the exclusive Racah matrices necessary for evaluating the arborescent knots \cite{MMMRSS} are yet unavailable for the representation $[3,1]$ \cite{MMMSsu}, while the inclusive Racah matrices in this representation are known \cite{MMMS31}. Hence, the integers in representations up to the fourth level can be constructed only for the knots that have 3-strand braid representations. The answers for these integers are summarized in the tables below.

Note that one may think the integrality of these numbers trivially follows from the integrality of the HOMFLY coefficients. In fact, this is completely non-trivial: if one considers just the HOMFLY polynomials rescaled with the framing factor $(-1)^{p|R|}q^{2p\varphi_R({[2]})}$, we see  that  these framed HOMFLY polynomials also obeys the integrality property. For example, one of the $p$ dependent coefficient, with this factor multiplied  $A^{-2}(q-q^{-1})^{9}$, is

{\footnotesize
\be
N_{[2],-2,6}={1\over 3832012800}\Big( 148p^{12}-2736p^{11}+79112p^{10}-831600p^9+10539474p^8-68756688p^7+436908296p^6-\\ -1721451600p^5+5409488128p^4 -11272637376p^3+15223732992p^2+2338875\cdot (-1)^p-11844403200p+3829673925\Big)\nn
\ee
}
\noindent
Note that the huge denominator $3832012800=12!2^3$ gets cancelled with the numerator for integer values of  $p$. Further we observe that the number of non-zero integers increases with increasing $|p|$ starting from large enough values of $p$, and the integers themselves celebrate some additional constraints, e.g.
\be
N_{[p]}(p)=(-1)^pN_{[1^p]}(p+1)
\ee
Looking at the tables below, one may note the two properties: all the numbers in each column have the same sign (it alternates with turning at some value) and the sum of all the coefficients in each row is equal to zero. The first property, though being correct very often still sometimes breaks: for instance, for the twist knots (which have maximal braid number at the given number of crossings) starting from knot $6_1$ already for Young diagrams of level 2. The second property follows from the fact that the unreduced HOMFLY polynomials are cancelled at $A=1$, and, hence, so do $f_R$ (\ref{ic}). From this latter formula and the fact that $\psi_R(\Delta)$ are symmetric group characters and, hence, are linearly independent it follows that
\be
\sum_n N_{{\bf Q},n,k}=0
\ee
at least, up to the level $|Q|=4$, where $^*p_\Delta$ are all independent.

\textbf{Knot} $8_{20}$:

\bigskip

 \begin{tabular}{cc}
$\mathbf{N}_{[ 1]}:$ &
\begin{tabular}{|c|cccc|}
\hline
&&&&\\
$ k \backslash n=$ & -5 & -3 & -1 & 1 \\
&&&&\\
\hline
&&&&\\
0 & 2 & -6 & 5 & -1 \\
&&&&\\
1 & 1 & -5 & 5 & -1 \\
&&&&\\
2 & 0 & -1 & 1 & 0 \\
&&&&\\
\hline
\end{tabular}
\end{tabular}

\hspace{-1cm}\begin{tabular}{cccc}
$\mathbf{N}_{[ 2]}:$ &
\begin{tabular}{|c|ccccccc|}
\hline
&&&&&&&\\
$ k \backslash n=$ & -10 & -8 & -6 & -4 & -2 & 0 & 2 \\
&&&&&&&\\
\hline
&&&&&&&\\
0 & 16 & -73 & 131 & -114 & 46 & -5 & -1 \\
&&&&&&&\\
1 & 50 & -231 & 400 & -319 & 111 & -10 & -1 \\
&&&&&&&\\
2 & 63 & -309 & 521 & -373 & 104 & -6 & 0 \\
&&&&&&&\\
3 & 37 & -212 & 359 & -231 & 48 & -1 & 0 \\
&&&&&&&\\
4 & 10 & -77 & 135 & -79 & 11 & 0 & 0 \\
&&&&&&&\\
5 & 1 & -14 & 26 & -14 & 1 & 0 & 0 \\
&&&&&&&\\
6 & 0 & -1 & 2 & -1 & 0 & 0 & 0 \\
&&&&&&&\\
\hline
\end{tabular}&
$\mathbf{N}_{[ 1, 1]}:$ &
\begin{tabular}{|c|cccccc|}
\hline
&&&&&&\\
$ k \backslash n=$ & -10 & -8 & -6 & -4 & -2 & 0 \\
&&&&&&\\
\hline
&&&&&&\\
0 & 25 & -115 & 210 & -190 & 85 & -15 \\
&&&&&&\\
1 & 95 & -440 & 775 & -645 & 250 & -35 \\
&&&&&&\\
2 & 155 & -743 & 1267 & -953 & 302 & -28 \\
&&&&&&\\
3 & 129 & -680 & 1148 & -781 & 193 & -9 \\
&&&&&&\\
4 & 56 & -354 & 607 & -377 & 69 & -1 \\
&&&&&&\\
5 & 12 & -104 & 185 & -106 & 13 & 0 \\
&&&&&&\\
6 & 1 & -16 & 30 & -16 & 1 & 0 \\
&&&&&&\\
7 & 0 & -1 & 2 & -1 & 0 & 0 \\
&&&&&&\\
\hline
\end{tabular}
\end{tabular}

{\footnotesize
 \begin{tabular}{cc}
$\mathbf{N}_{[ 3]}:$ &
\begin{tabular}{|c|cccccccccc|}
\hline
&&&&&&&&&&\\
$ k \backslash n=$ & -15 & -13 & -11 & -9 & -7 & -5 & -3 & -1 & 1 & 3 \\
&&&&&&&&&&\\
\hline
&&&&&&&&&&\\
0 & 352 & -2125 & 5468 & -7791 & 6673 & -3470 & 1022 & -111 & -27 & 9 \\
&&&&&&&&&&\\
1 & 3256 & -18695 & 44944 & -58584 & 44782 & -20245 & 5097 & -473 & -123 & 41 \\
&&&&&&&&&&\\
2 & 14770 & -81370 & 183559 & -218724 & 147871 & -56664 & 11559 & -853 & -209 & 61 \\
&&&&&&&&&&\\
3 & 41511 & -222579 & 475465 & -520438 & 310866 & -99510 & 15655 & -842 & -165 & 37 \\
&&&&&&&&&&\\
4 & 77904 & -414115 & 847003 & -857240 & 453256 & -120239 & 13971 & -484 & -66 & 10 \\
&&&&&&&&&&\\
5 & 101052 & -543578 & 1076296 & -1013191 & 474743 & -103702 & 8552 & -160 & -13 & 1 \\
&&&&&&&&&&\\
6 & 92372 & -514010 & 995015 & -874977 & 362651 & -64643 & 3621 & -28 & -1 & 0 \\
&&&&&&&&&&\\
7 & 60098 & -354425 & 676451 & -557025 & 202945 & -29084 & 1042 & -2 & 0 & 0 \\
&&&&&&&&&&\\
8 & 27855 & -179063 & 339359 & -261883 & 82860 & -9322 & 194 & 0 & 0 & 0 \\
&&&&&&&&&&\\
9 & 9107 & -66077 & 125094 & -90413 & 24338 & -2070 & 21 & 0 & 0 & 0 \\
&&&&&&&&&&\\
10 & 2048 & -17576 & 33405 & -22574 & 4998 & -302 & 1 & 0 & 0 & 0 \\
&&&&&&&&&&\\
11 & 301 & -3277 & 6279 & -3957 & 680 & -26 & 0 & 0 & 0 & 0 \\
&&&&&&&&&&\\
12 & 26 & -406 & 787 & -461 & 55 & -1 & 0 & 0 & 0 & 0 \\
&&&&&&&&&&\\
13 & 1 & -30 & 59 & -32 & 2 & 0 & 0 & 0 & 0 & 0 \\
&&&&&&&&&&\\
14 & 0 & -1 & 2 & -1 & 0 & 0 & 0 & 0 & 0 & 0 \\
&&&&&&&&&&\\
\hline
\end{tabular}
\end{tabular}}

\hspace{-1.5cm}\begin{tabular}{cc}
$\mathbf{N}_{[ 2, 1]}:$ &
\begin{tabular}{|c|cccccccccc|}
\hline
&&&&&&&&&&\\
$ k \backslash n=$ & -15 & -13 & -11 & -9 & -7 & -5 & -3 & -1 & 1 & 3 \\
&&&&&&&&&&\\
\hline
&&&&&&&&&&\\
0 & 1096 & -6812 & 18055 & -26511 & 23427 & -12647 & 4021 & -661 & 25 & 7 \\
&&&&&&&&&&\\
1 & 11740 & -69190 & 171467 & -231418 & 183976 & -86971 & 23475 & -3220 & 118 & 23 \\
&&&&&&&&&&\\
2 & 62734 & -352952 & 820140 & -1016046 & 721600 & -293594 & 64694 & -6802 & 204 & 22 \\
&&&&&&&&&&\\
3 & 211059 & -1148176 & 2521968 & -2880042 & 1826520 & -634085 & 110761 & -8177 & 164 & 8 \\
&&&&&&&&&&\\
4 & 482979 & -2581587 & 5415098 & -5735572 & 3255233 & -960113 & 130062 & -6167 & 66 & 1 \\
&&&&&&&&&&\\
5 & 780717 & -4170723 & 8440154 & -8341780 & 4245634 & -1060226 & 109232 & -3021 & 13 & 0 \\
&&&&&&&&&&\\
6 & 912409 & -4956911 & 9766956 & -9052011 & 4133483 & -869612 & 66643 & -958 & 1 & 0 \\
&&&&&&&&&&\\
7 & 782632 & -4402394 & 8512833 & -7424291 & 3035652 & -533774 & 29531 & -189 & 0 & 0 \\
&&&&&&&&&&\\
8 & 496928 & -2950758 & 5637051 & -4636024 & 1688623 & -245177 & 9378 & -21 & 0 & 0 \\
&&&&&&&&&&\\
9 & 234028 & -1499349 & 2845722 & -2209021 & 710203 & -83655 & 2073 & -1 & 0 & 0 \\
&&&&&&&&&&\\
10 & 81306 & -576851 & 1092898 & -800716 & 223914 & -20853 & 302 & 0 & 0 & 0 \\
&&&&&&&&&&\\
11 & 20526 & -166688 & 316505 & -218681 & 51995 & -3683 & 26 & 0 & 0 & 0 \\
&&&&&&&&&&\\
12 & 3656 & -35556 & 67891 & -44172 & 8616 & -436 & 1 & 0 & 0 & 0 \\
&&&&&&&&&&\\
13 & 435 & -5426 & 10448 & -6389 & 963 & -31 & 0 & 0 & 0 & 0 \\
&&&&&&&&&&\\
14 & 31 & -560 & 1090 & -625 & 65 & -1 & 0 & 0 & 0 & 0 \\
&&&&&&&&&&\\
15 & 1 & -35 & 69 & -37 & 2 & 0 & 0 & 0 & 0 & 0 \\
&&&&&&&&&&\\
16 & 0 & -1 & 2 & -1 & 0 & 0 & 0 & 0 & 0 & 0 \\
&&&&&&&&&&\\
\hline
\end{tabular}
\end{tabular}

\hspace{-2cm}\begin{tabular}{cc}
$\mathbf{N}_{[ 1, 1, 1]}:$ &
\begin{tabular}{|c|ccccccccc|}
\hline
&&&&&&&&&\\
$ k \backslash n=$ & -15 & -13 & -11 & -9 & -7 & -5 & -3 & -1 & 1 \\
&&&&&&&&&\\
\hline
&&&&&&&&&\\
0 & 817 & -5202 & 14122 & -21247 & 19265 & -10712 & 3552 & -647 & 52 \\
&&&&&&&&&\\
1 & 9896 & -59591 & 151295 & -209830 & 172040 & -84226 & 23656 & -3445 & 205 \\
&&&&&&&&&\\
2 & 60278 & -345235 & 821309 & -1048249 & 772553 & -328797 & 76150 & -8309 & 300 \\
&&&&&&&&&\\
3 & 232831 & -1283378 & 2881690 & -3396607 & 2249185 & -827003 & 155091 & -12020 & 211 \\
&&&&&&&&&\\
4 & 616432 & -3318766 & 7100687 & -7772468 & 4632525 & -1466503 & 219517 & -11501 & 77 \\
&&&&&&&&&\\
5 & 1162736 & -6209959 & 12783754 & -13070245 & 7025091 & -1908577 & 224707 & -7521 & 14 \\
&&&&&&&&&\\
6 & 1600769 & -8613912 & 17213476 & -16520487 & 8014837 & -1860533 & 169213 & -3364 & 1 \\
&&&&&&&&&\\
7 & 1634944 & -9004917 & 17603582 & -15919558 & 6965468 & -1372676 & 94164 & -1007 & 0 \\
&&&&&&&&&\\
8 & 1251705 & -7174091 & 13810549 & -11798458 & 4641249 & -769287 & 38525 & -192 & 0 \\
&&&&&&&&&\\
9 & 721849 & -4384030 & 8357680 & -6754633 & 2374468 & -326722 & 11409 & -21 & 0 \\
&&&&&&&&&\\
10 & 313286 & -2058624 & 3905215 & -2987163 & 929118 & -104204 & 2373 & -1 & 0 \\
&&&&&&&&&\\
11 & 101531 & -740262 & 1403124 & -1015440 & 275229 & -24510 & 328 & 0 & 0 \\
&&&&&&&&&\\
12 & 24156 & -201838 & 383609 & -262392 & 60556 & -4118 & 27 & 0 & 0 \\
&&&&&&&&&\\
13 & 4090 & -40952 & 78280 & -50529 & 9577 & -467 & 1 & 0 & 0 \\
&&&&&&&&&\\
14 & 466 & -5985 & 11536 & -7013 & 1028 & -32 & 0 & 0 & 0 \\
&&&&&&&&&\\
15 & 32 & -595 & 1159 & -662 & 67 & -1 & 0 & 0 & 0 \\
&&&&&&&&&\\
16 & 1 & -36 & 71 & -38 & 2 & 0 & 0 & 0 & 0 \\
&&&&&&&&&\\
17 & 0 & -1 & 2 & -1 & 0 & 0 & 0 & 0 & 0 \\
&&&&&&&&&\\
\hline
\end{tabular}
\end{tabular}

\newpage

$\mathbf{N}_{[ 4]}:$

\bigskip

\hspace{-2cm}{\tiny
\begin{tabular}{|c|ccccccccccccc|}
\hline
&&&&&&&&&&&&&\\
$ k \backslash n=$ & -20 & -18 & -16 & -14 & -12 & -10 & -8 & -6 & -4 & -2 & 0 & 2 & 4 \\
&&&&&&&&&&&&&\\
\hline
&&&&&&&&&&&&&\\
0 & 11440 & -87173 & 293893 & -576270 & 726572 & -614639 & 352840 & -135087 & 31946 & -3116 & -645 & 269 & -30 \\
&&&&&&&&&&&&&\\
1 & 228250 & -1635276 & 5137191 & -9286702 & 10657519 & -8081601 & 4086664 & -1355188 & 275713 & -23835 & -4413 & 1934 & -256 \\
&&&&&&&&&&&&&\\
2 & 2386083 & -16136564 & 47369508 & -79062965 & 82549940 & -55906878 & 24662653 & -6936301 & 1167700 & -85260 & -13024 & 6054 & -946 \\
&&&&&&&&&&&&&\\
3 & 16661172 & -107057436 & 295234625 & -456862001 & 435207944 & -263432562 & 101033685 & -23791571 & 3205258 & -186183 & -21779 & 10693 & -1845 \\
&&&&&&&&&&&&&\\
4 & 84507887 & -519907050 & 1355251916 & -1954018736 & 1704701161 & -924387333 & 308130762 & -60328166 & 6336339 & -273569 & -22825 & 11672 & -2058 \\
&&&&&&&&&&&&&\\
5 & 324218115 & -1925277128 & 4773608162 & -6443894762 & 5167017105 & -2515811031 & 729145727 & -118202574 & 9488612 & -283394 & -15660 & 8205 & -1377 \\
&&&&&&&&&&&&&\\
6 & 964060168 & -5570786966 & 13216800689 & -16777389869 & 12401705555 & -5430400831 & 1368154785 & -182973872 & 11046274 & -212013 & -7130 & 3771 & -561 \\
&&&&&&&&&&&&&\\
7 & 2260162822 & -12811019508 & 29245594272 & -35043455568 & 23932680078 & -9429839738 & 2062360089 & -226492258 & 10126690 & -115736 & -2131 & 1124 & -136 \\
&&&&&&&&&&&&&\\
8 & 4233247530 & -23722411643 & 52372794915 & -59428427145 & 37551473309 & -13307457043 & 2519220675 & -225742273 & 7347950 & -46065 & -401 & 209 & -18 \\
&&&&&&&&&&&&&\\
9 & 6401577363 & -35742184785 & 76665320902 & -82599999575 & 48322505041 & -15378284326 & 2508707131 & -181847621 & 4219095 & -13203 & -43 & 22 & -1 \\
&&&&&&&&&&&&&\\
10 & 7882325987 & -44189952927 & 92475676765 & -94805633387 & 51345839966 & -14636437277 & 2044843043 & -118567832 & 1908313 & -2650 & -2 & 1 & 0 \\
&&&&&&&&&&&&&\\
11 & 7955332198 & -45135236322 & 92502848394 & -90392118364 & 45284233622 & -11520499158 & 1367294612 & -62528465 & 673836 & -353 & 0 & 0 & 0 \\
&&&&&&&&&&&&&\\
12 & 6613694775 & -38282967105 & 77104986495 & -71912996672 & 33272091742 & -7518694886 & 750290984 & -26588381 & 183076 & -28 & 0 & 0 & 0 \\
&&&&&&&&&&&&&\\
13 & 4544009056 & -27063745212 & 53736998578 & -47883973931 & 20411212868 & -4072813151 & 337335831 & -9061459 & 37421 & -1 & 0 & 0 & 0 \\
&&&&&&&&&&&&&\\
14 & 2584289462 & -15981116400 & 31372940707 & -26729420170 & 10461927572 & -1829952056 & 123776261 & -2450930 & 5554 & 0 & 0 & 0 & 0 \\
&&&&&&&&&&&&&\\
15 & 1216390788 & -7887415145 & 15349485952 & -12510461549 & 4475947275 & -680239806 & 36810141 & -518220 & 564 & 0 & 0 & 0 & 0 \\
&&&&&&&&&&&&&\\
16 & 472819775 & -3250240468 & 6285479701 & -4902402599 & 1593840148 & -208191240 & 8778345 & -83697 & 35 & 0 & 0 & 0 & 0 \\
&&&&&&&&&&&&&\\
17 & 151097941 & -1114995248 & 2147461941 & -1603116813 & 469971253 & -52061194 & 1652078 & -9959 & 1 & 0 & 0 & 0 & 0 \\
&&&&&&&&&&&&&\\
18 & 39405913 & -316778506 & 608869085 & -435066095 & 113847320 & -10516465 & 239570 & -822 & 0 & 0 & 0 & 0 & 0 \\
&&&&&&&&&&&&&\\
19 & 8294017 & -73942242 & 142098694 & -97183015 & 22394545 & -1687755 & 25798 & -42 & 0 & 0 & 0 & 0 & 0 \\
&&&&&&&&&&&&&\\
20 & 1385933 & -14015257 & 26975820 & -17655709 & 3517260 & -209987 & 1941 & -1 & 0 & 0 & 0 & 0 & 0 \\
&&&&&&&&&&&&&\\
21 & 179446 & -2120970 & 4095152 & -2564506 & 430297 & -19510 & 91 & 0 & 0 & 0 & 0 & 0 & 0 \\
&&&&&&&&&&&&&\\
22 & 17344 & -249997 & 484911 & -290476 & 39489 & -1273 & 2 & 0 & 0 & 0 & 0 & 0 & 0 \\
&&&&&&&&&&&&&\\
23 & 1177 & -22102 & 43125 & -24704 & 2556 & -52 & 0 & 0 & 0 & 0 & 0 & 0 & 0 \\
&&&&&&&&&&&&&\\
24 & 50 & -1378 & 2708 & -1483 & 104 & -1 & 0 & 0 & 0 & 0 & 0 & 0 & 0 \\
&&&&&&&&&&&&&\\
25 & 1 & -54 & 107 & -56 & 2 & 0 & 0 & 0 & 0 & 0 & 0 & 0 & 0 \\
&&&&&&&&&&&&&\\
26 & 0 & -1 & 2 & -1 & 0 & 0 & 0 & 0 & 0 & 0 & 0 & 0 & 0 \\
&&&&&&&&&&&&&\\
\hline
\end{tabular}
}

\newpage
\begin{landscape}
$\mathbf{N}_{[ 2, 2]}:$

\bigskip

{\tiny
\begin{tabular}{|c|ccccccccccccc|}
\hline
&&&&&&&&&&&&&\\
$ k \backslash n=$ & -20 & -18 & -16 & -14 & -12 & -10 & -8 & -6 & -4 & -2 & 0 & 2 & 4 \\
&&&&&&&&&&&&&\\
\hline
&&&&&&&&&&&&&\\
0 & 45142 & -356604 & 1248004 & -2544260 & 3341462 & -2951474 & 1776512 & -721950 & 191238 & -30194 & 2020 & 130 & -26 \\
&&&&&&&&&&&&&\\
1 & 1061746 & -7886896 & 25774891 & -48658404 & 58570857 & -46814370 & 25091891 & -8900632 & 2010383 & -265010 & 14986 & 704 & -146 \\
&&&&&&&&&&&&&\\
2 & 13219093 & -92644735 & 283364735 & -495881207 & 546856462 & -394544930 & 187217848 & -57276214 & 10767963 & -1129629 & 49369 & 1575 & -330 \\
&&&&&&&&&&&&&\\
3 & 110978071 & -738218369 & 2123395303 & -3457325739 & 3501914639 & -2283015929 & 957875396 & -251314127 & 38722459 & -3108818 & 95608 & 1880 & -374 \\
&&&&&&&&&&&&&\\
4 & 683626377 & -4346342959 & 11823796029 & -17993793212 & 16798542200 & -9918676313 & 3682017558 & -826512903 & 103384154 & -6163505 & 121495 & 1310 & -231 \\
&&&&&&&&&&&&&\\
5 & 3220746083 & -19713056578 & 51012920266 & -72893664160 & 62941377502 & -33738218362 & 11091817656 & -2127319330 & 214560357 & -9270796 & 106893 & 548 & -79 \\
&&&&&&&&&&&&&\\
6 & 11902797889 & -70635231409 & 174837831840 & -235577769444 & 188722760807 & -92013045366 & 26804397519 & -4385326880 & 354369716 & -10851594 & 66801 & 135 & -14 \\
&&&&&&&&&&&&&\\
7 & 35131751461 & -203521505109 & 484315599846 & -617676693690 & 460269219324 & -204383089677 & 52746115436 & -7343388151 & 471962933 & -10002267 & 29877 & 18 & -1 \\
&&&&&&&&&&&&&\\
8 & 83977371481 & -478017884947 & 1098713484889 & -1330695425149 & 924190522239 & -374008482052 & 85421742814 & -10084961299 & 510913581 & -7291050 & 9492 & 1 & 0 \\
&&&&&&&&&&&&&\\
9 & 164416998734 & -925383009601 & 2063046363926 & -2379476149635 & 1542491103249 & -568872395551 & 114766633622 & -11437048121 & 451701955 & -4200668 & 2090 & 0 & 0 \\
&&&&&&&&&&&&&\\
10 & 266136440145 & -1490091389662 & 3234417542766 & -3560978652096 & 2156541392942 & -724294109728 & 128707348520 & -10763632855 & 326963860 & -1904195 & 303 & 0 & 0 \\
&&&&&&&&&&&&&\\
11 & 358904242033 & -2011015976626 & 4264679863987 & -4490663780569 & 2541586916172 & -776296340555 & 121046412016 & -8434521195 & 193857944 & -673233 & 26 & 0 & 0 \\
&&&&&&&&&&&&&\\
12 & 405783962986 & -2288990129732 & 4757265150324 & -4798780206020 & 2537845729254 & -703490616645 & 95785780542 & -5513478795 & 93991109 & -183024 & 1 & 0 & 0 \\
&&&&&&&&&&&&&\\
13 & 386560168725 & -2208448533446 & 4511053037295 & -4364839402142 & 2155514108908 & -540773914262 & 63904984526 & -3007528055 & 37115870 & -37419 & 0 & 0 & 0 \\
&&&&&&&&&&&&&\\
14 & 311445592580 & -1813201702273 & 3649582846272 & -3390820641320 & 1561745471082 & -353369057566 & 35972944947 & -1367302839 & 11854671 & -5554 & 0 & 0 & 0 \\
&&&&&&&&&&&&&\\
15 & 212773261251 & -1270437768319 & 2525765933165 & -2255190702268 & 967019262973 & -196491288944 & 17074742058 & -516469165 & 3029813 & -564 & 0 & 0 & 0 \\
&&&&&&&&&&&&&\\
16 & 123435129401 & -760994699619 & 1497674084990 & -1285908719153 & 512097029310 & -92961530399 & 6819326161 & -161230450 & 609794 & -35 & 0 & 0 & 0 \\
&&&&&&&&&&&&&\\
17 & 60818637949 & -389973798110 & 761283513563 & -628846628374 & 231845177029 & -37368435345 & 2282702661 & -41263734 & 94362 & -1 & 0 & 0 & 0 \\
&&&&&&&&&&&&&\\
18 & 25423341446 & -170894213960 & 331532675036 & -263555527248 & 89592696958 & -12727026265 & 636599926 & -8556713 & 10820 & 0 & 0 & 0 & 0 \\
&&&&&&&&&&&&&\\
19 & 8993174943 & -63934192335 & 123473129260 & -94483488337 & 29462284035 & -3656119893 & 146625003 & -1413541 & 865 & 0 & 0 & 0 & 0 \\
&&&&&&&&&&&&&\\
20 & 2680538218 & -20357277502 & 39200955717 & -28877797256 & 8206333118 & -880120523 & 27549660 & -181475 & 43 & 0 & 0 & 0 & 0 \\
&&&&&&&&&&&&&\\
21 & 668955054 & -5490782054 & 10558350417 & -7487841536 & 1923099476 & -175913501 & 4149580 & -17437 & 1 & 0 & 0 & 0 & 0 \\
&&&&&&&&&&&&&\\
22 & 138520892 & -1246055050 & 2395991231 & -1635750854 & 375629940 & -28823523 & 488543 & -1179 & 0 & 0 & 0 & 0 & 0 \\
&&&&&&&&&&&&&\\
23 & 23501434 & -235692966 & 453774067 & -298191597 & 60368864 & -3803029 & 43277 & -50 & 0 & 0 & 0 & 0 & 0 \\
&&&&&&&&&&&&&\\
24 & 3209465 & -36680505 & 70793497 & -44771764 & 7840441 & -393844 & 2711 & -1 & 0 & 0 & 0 & 0 & 0 \\
&&&&&&&&&&&&&\\
25 & 343919 & -4612864 & 8934653 & -5436982 & 801973 & -30806 & 107 & 0 & 0 & 0 & 0 & 0 & 0 \\
&&&&&&&&&&&&&\\
26 & 27830 & -456835 & 888927 & -520375 & 62161 & -1710 & 2 & 0 & 0 & 0 & 0 & 0 & 0 \\
&&&&&&&&&&&&&\\
27 & 1598 & -34280 & 67076 & -37764 & 3430 & -60 & 0 & 0 & 0 & 0 & 0 & 0 & 0 \\
&&&&&&&&&&&&&\\
28 & 58 & -1831 & 3606 & -1952 & 120 & -1 & 0 & 0 & 0 & 0 & 0 & 0 & 0 \\
&&&&&&&&&&&&&\\
29 & 1 & -62 & 123 & -64 & 2 & 0 & 0 & 0 & 0 & 0 & 0 & 0 & 0 \\
&&&&&&&&&&&&&\\
30 & 0 & -1 & 2 & -1 & 0 & 0 & 0 & 0 & 0 & 0 & 0 & 0 & 0 \\
&&&&&&&&&&&&&\\
\hline
\end{tabular}
}

\newpage

$\mathbf{N}_{[ 3, 1]}:$

\bigskip

{\tiny
\begin{tabular}{|c|ccccccccccccc|}
\hline
&&&&&&&&&&&&&\\
$ k \backslash n=$ & -20 & -18 & -16 & -14 & -12 & -10 & -8 & -6 & -4 & -2 & 0 & 2 & 4 \\
&&&&&&&&&&&&&\\
\hline
&&&&&&&&&&&&&\\
0 & 54046 & -421913 & 1458902 & -2937624 & 3808680 & -3318671 & 1968098 & -785583 & 202064 & -29280 & 920 & 415 & -54 \\
&&&&&&&&&&&&&\\
1 & 1207930 & -8866816 & 28610436 & -53272255 & 63169584 & -49668944 & 26149099 & -9092365 & 2000388 & -246716 & 7507 & 2500 & -348 \\
&&&&&&&&&&&&&\\
2 & 14241824 & -98651137 & 297765825 & -513306085 & 556454090 & -393694624 & 182719619 & -54537515 & 9968959 & -992881 & 26593 & 6264 & -932 \\
&&&&&&&&&&&&&\\
3 & 112850652 & -742218311 & 2106166482 & -3374445171 & 3353073386 & -2136652475 & 872510806 & -221807292 & 32987994 & -2526777 & 53538 & 8456 & -1288 \\
&&&&&&&&&&&&&\\
4 & 653965466 & -4113310890 & 11037788852 & -16513730121 & 15095228135 & -8683571028 & 3121017399 & -673245823 & 80317835 & -4533288 & 67664 & 6800 & -1001 \\
&&&&&&&&&&&&&\\
5 & 2888733279 & -17506113346 & 44687098358 & -62724026193 & 52933567411 & -27554780281 & 8724052977 & -1593494579 & 150952041 & -6049005 & 56385 & 3408 & -455 \\
&&&&&&&&&&&&&\\
6 & 9975276084 & -58675127063 & 143285992377 & -189504490234 & 148106635013 & -69899466805 & 19503109153 & -3010288126 & 224500982 & -6174069 & 31735 & 1073 & -120 \\
&&&&&&&&&&&&&\\
7 & 27413969447 & -157632856262 & 370192623268 & -463092032177 & 336031568249 & -143958430444 & 35384632338 & -4602450447 & 267852157 & -4888444 & 12126 & 206 & -17 \\
&&&&&&&&&&&&&\\
8 & 60794764793 & -344088847169 & 780828884205 & -926938058004 & 625678990120 & -243423323913 & 52639550626 & -5747305587 & 258360292 & -3018483 & 3099 & 22 & -1 \\
&&&&&&&&&&&&&\\
9 & 110022802044 & -617025495856 & 1358843153531 & -1535082506469 & 965128511825 & -340890212578 & 64701897506 & -5898989939 & 202291605 & -1452177 & 507 & 1 & 0 \\
&&&&&&&&&&&&&\\
10 & 163995554522 & -917250958025 & 1968031062413 & -2120675962227 & 1242763450545 & -398084069590 & 66090861594 & -4998150672 & 128752218 & -540826 & 48 & 0 & 0 \\
&&&&&&&&&&&&&\\
11 & 202863894726 & -1138898560506 & 2389144523104 & -2460370558556 & 1344113111105 & -389755815577 & 56342255075 & -3505241905 & 66546492 & -153960 & 2 & 0 & 0 \\
&&&&&&&&&&&&&\\
12 & 209533734845 & -1188374089005 & 2445223534552 & -2410319996901 & 1227007686986 & -321262829015 & 40201070394 & -2036913022 & 27833969 & -32803 & 0 & 0 & 0 \\
&&&&&&&&&&&&&\\
13 & 181570438382 & -1047138101394 & 2119566035105 & -2002402805737 & 948918733102 & -223586809674 & 24043357815 & -980203313 & 9360768 & -5054 & 0 & 0 & 0 \\
&&&&&&&&&&&&&\\
14 & 132459443162 & -782044704685 & 1561389673170 & -1415147162422 & 623288956096 & -131612894137 & 12053834213 & -389650341 & 2505475 & -531 & 0 & 0 & 0 \\
&&&&&&&&&&&&&\\
15 & 81530208844 & -496280477552 & 979725736282 & -852553990370 & 348206309830 & -65558758888 & 5057797190 & -127350800 & 525498 & -34 & 0 & 0 & 0 \\
&&&&&&&&&&&&&\\
16 & 42377740331 & -267964462462 & 524233827595 & -438257130491 & 165484567211 & -27610978919 & 1770317014 & -33964645 & 84367 & -1 & 0 & 0 & 0 \\
&&&&&&&&&&&&&\\
17 & 18591995559 & -123124330894 & 239197147867 & -192192614652 & 66830581769 & -9809519063 & 514038540 & -7309123 & 9997 & 0 & 0 & 0 & 0 \\
&&&&&&&&&&&&&\\
18 & 6870751800 & -48086661572 & 92944396253 & -71797978690 & 22875962548 & -2928023746 & 122800998 & -1248414 & 823 & 0 & 0 & 0 & 0 \\
&&&&&&&&&&&&&\\
19 & 2130717160 & -15921473005 & 30670831701 & -22782692662 & 6608716033 & -729783904 & 23849804 & -165169 & 42 & 0 & 0 & 0 & 0 \\
&&&&&&&&&&&&&\\
20 & 551206991 & -4449819754 & 8557099836 & -6112760303 & 1601134345 & -150546606 & 3701797 & -16307 & 1 & 0 & 0 & 0 & 0 \\
&&&&&&&&&&&&&\\
21 & 117927509 & -1043083270 & 2005345733 & -1377645739 & 322395428 & -25386405 & 447874 & -1130 & 0 & 0 & 0 & 0 & 0 \\
&&&&&&&&&&&&&\\
22 & 20610727 & -203221777 & 391130409 & -258395591 & 53274001 & -3438391 & 40671 & -49 & 0 & 0 & 0 & 0 & 0 \\
&&&&&&&&&&&&&\\
23 & 2891884 & -32493291 & 62686783 & -39820710 & 7097419 & -364690 & 2606 & -1 & 0 & 0 & 0 & 0 & 0 \\
&&&&&&&&&&&&&\\
24 & 317631 & -4188592 & 8109422 & -4952537 & 743126 & -29155 & 105 & 0 & 0 & 0 & 0 & 0 & 0 \\
&&&&&&&&&&&&&\\
25 & 26289 & -424326 & 825338 & -484501 & 58849 & -1651 & 2 & 0 & 0 & 0 & 0 & 0 & 0 \\
&&&&&&&&&&&&&\\
26 & 1541 & -32510 & 63591 & -35875 & 3312 & -59 & 0 & 0 & 0 & 0 & 0 & 0 & 0 \\
&&&&&&&&&&&&&\\
27 & 57 & -1770 & 3485 & -1889 & 118 & -1 & 0 & 0 & 0 & 0 & 0 & 0 & 0 \\
&&&&&&&&&&&&&\\
28 & 1 & -61 & 121 & -63 & 2 & 0 & 0 & 0 & 0 & 0 & 0 & 0 & 0 \\
&&&&&&&&&&&&&\\
29 & 0 & -1 & 2 & -1 & 0 & 0 & 0 & 0 & 0 & 0 & 0 & 0 & 0 \\
&&&&&&&&&&&&&\\
\hline
\end{tabular}
}

\newpage

$\mathbf{N}_{[ 2, 1, 1]}:$

\bigskip

{\tiny
\begin{tabular}{|c|ccccccccccccc|}
\hline
&&&&&&&&&&&&&\\
$ k \backslash n=$ & -20 & -18 & -16 & -14 & -12 & -10 & -8 & -6 & -4 & -2 & 0 & 2 & 4 \\
&&&&&&&&&&&&&\\
\hline
&&&&&&&&&&&&&\\
0 & 82226 & -655421 & 2315601 & -4768098 & 6328599 & -5653333 & 3444750 & -1419723 & 383294 & -62978 & 5093 & 17 & -27 \\
&&&&&&&&&&&&&\\
1 & 2028650 & -15204532 & 50179191 & -95758205 & 116651501 & -94485833 & 51405282 & -18548132 & 4277813 & -584321 & 38779 & -69 & -124 \\
&&&&&&&&&&&&&\\
2 & 26498440 & -187350248 & 578823419 & -1024679559 & 1145145541 & -839062818 & 405427370 & -126727185 & 24457479 & -2665562 & 133830 & -498 & -209 \\
&&&&&&&&&&&&&\\
3 & 233347245 & -1565447193 & 4548843934 & -7496837137 & 7704425860 & -5111284629 & 2190726176 & -590195832 & 94065559 & -7920853 & 277991 & -956 & -165 \\
&&&&&&&&&&&&&\\
4 & 1507954814 & -9664436055 & 26558995186 & -40929741986 & 38809049674 & -23361113876 & 8885693743 & -2058362496 & 268617640 & -17043660 & 387951 & -869 & -66 \\
&&&&&&&&&&&&&\\
5 & 7457193738 & -45979882379 & 120177405182 & -173955538607 & 152695842572 & -83583875475 & 28235741941 & -5615477154 & 596110616 & -27903222 & 383230 & -429 & -13 \\
&&&&&&&&&&&&&\\
6 & 28953907377 & -172943961905 & 432241801187 & -590126024957 & 481008054221 & -239881644844 & 72003333217 & -12273244597 & 1053142525 & -35636775 & 274670 & -118 & -1 \\
&&&&&&&&&&&&&\\
7 & 89884104520 & -523567063468 & 1257592724079 & -1625517784143 & 1233501689839 & -561205891237 & 149658769719 & -21813116596 & 1502387244 & -35964061 & 144121 & -17 & 0 \\
&&&&&&&&&&&&&\\
8 & 226269108658 & -1293486377314 & 2999627442970 & -3682825729662 & 2607163010053 & -1082986025538 & 256372018615 & -31850983340 & 1746335725 & -28855399 & 55233 & -1 & 0 \\
&&&&&&&&&&&&&\\
9 & 467205567433 & -2637141435217 & 5929009498223 & -6934041621715 & 4586535710544 & -1739735704517 & 365019802865 & -38497118179 & 1663725933 & -18440624 & 15254 & 0 & 0 \\
&&&&&&&&&&&&&\\
10 & 798830764924 & -4478406223815 & 9798267944342 & -10941736201568 & 6769351924270 & -2343670711258 & 434797827759 & -38729796224 & 1303841022 & -9372403 & 2951 & 0 & 0 \\
&&&&&&&&&&&&&\\
11 & 1139985271543 & -6384240759439 & 13639312613495 & -14572599444620 & 8437189046119 & -2663427324899 & 435504330724 & -32561954543 & 841991174 & -3769933 & 379 & 0 & 0 \\
&&&&&&&&&&&&&\\
12 & 1366722396147 & -7689612276847 & 16091251780551 & -16476684402749 & 8928116418760 & -2565505362470 & 368202683338 & -22938044558 & 447997301 & -1189502 & 29 & 0 & 0 \\
&&&&&&&&&&&&&\\
13 & 1383894030779 & -7867197628677 & 16170858227109 & -15890895606776 & 8055469323002 & -2102225549823 & 263457376308 & -13555859647 & 195977951 & -290227 & 1 & 0 & 0 \\
&&&&&&&&&&&&&\\
14 & 1188437503303 & -6866008935580 & 13898411902941 & -13122154163682 & 6217461566062 & -1469268192825 & 159769985154 & -6719786231 & 70174408 & -53550 & 0 & 0 & 0 \\
&&&&&&&&&&&&&\\
15 & 868235167489 & -5128234529678 & 10247154909485 & -9303838757754 & 4114596950171 & -877284165510 & 82139373931 & -2789359534 & 20418612 & -7212 & 0 & 0 & 0 \\
&&&&&&&&&&&&&\\
16 & 540700909123 & -3285457404714 & 6494554360540 & -5674659729236 & 2337825783692 & -447764901222 & 35762403527 & -966195870 & 4774828 & -668 & 0 & 0 & 0 \\
&&&&&&&&&&&&&\\
17 & 287299047689 & -1807791763055 & 3542327477812 & -2980255989623 & 1140801978628 & -195257317371 & 13153395604 & -277712529 & 882883 & -38 & 0 & 0 & 0 \\
&&&&&&&&&&&&&\\
18 & 130217910487 & -854569778418 & 1662937142418 & -1347715231428 & 477747305355 & -72621490394 & 4069701441 & -65685459 & 125999 & -1 & 0 & 0 & 0 \\
&&&&&&&&&&&&&\\
19 & 50272149115 & -346775317960 & 671281231115 & -524223299597 & 171370833030 & -22965602040 & 1052625131 & -12632167 & 13373 & 0 & 0 & 0 & 0 \\
&&&&&&&&&&&&&\\
20 & 16483582606 & -120556205087 & 232518896575 & -175004119802 & 52480149044 & -6145934856 & 225572186 & -1941659 & 993 & 0 & 0 & 0 & 0 \\
&&&&&&&&&&&&&\\
21 & 4569475871 & -35786540394 & 68870661235 & -49963676125 & 13653236118 & -1382474621 & 39550526 & -232656 & 46 & 0 & 0 & 0 & 0 \\
&&&&&&&&&&&&&\\
22 & 1063907003 & -9025725427 & 17355193701 & -12136698507 & 2996715295 & -258945679 & 5574538 & -20925 & 1 & 0 & 0 & 0 & 0 \\
&&&&&&&&&&&&&\\
23 & 206133310 & -1920640657 & 3694626649 & -2490504935 & 549639113 & -39867920 & 615768 & -1328 & 0 & 0 & 0 & 0 & 0 \\
&&&&&&&&&&&&&\\
24 & 32812198 & -341545889 & 658045136 & -427554352 & 83147061 & -4955406 & 51305 & -53 & 0 & 0 & 0 & 0 & 0 \\
&&&&&&&&&&&&&\\
25 & 4214933 & -50094771 & 96772118 & -60598209 & 10187511 & -484611 & 3030 & -1 & 0 & 0 & 0 & 0 & 0 \\
&&&&&&&&&&&&&\\
26 & 425868 & -5950859 & 11537843 & -6962232 & 985144 & -35877 & 113 & 0 & 0 & 0 & 0 & 0 & 0 \\
&&&&&&&&&&&&&\\
27 & 32567 & -557905 & 1086670 & -631778 & 72333 & -1889 & 2 & 0 & 0 & 0 & 0 & 0 & 0 \\
&&&&&&&&&&&&&\\
28 & 1771 & -39712 & 77773 & -43557 & 3788 & -63 & 0 & 0 & 0 & 0 & 0 & 0 & 0 \\
&&&&&&&&&&&&&\\
29 & 61 & -2016 & 3973 & -2143 & 126 & -1 & 0 & 0 & 0 & 0 & 0 & 0 & 0 \\
&&&&&&&&&&&&&\\
30 & 1 & -65 & 129 & -67 & 2 & 0 & 0 & 0 & 0 & 0 & 0 & 0 & 0 \\
&&&&&&&&&&&&&\\
31 & 0 & -1 & 2 & -1 & 0 & 0 & 0 & 0 & 0 & 0 & 0 & 0 & 0 \\
&&&&&&&&&&&&&\\
\hline
\end{tabular}
}

\newpage

$\mathbf{N}_{[ 1, 1, 1, 1]}:$

\bigskip

{\tiny
\begin{tabular}{|c|ccccccccccccc|}
\hline
&&&&&&&&&&&&&\\
$ k \backslash n=$ & -20 & -18 & -16 & -14 & -12 & -10 & -8 & -6 & -4 & -2 & 0 & 2 & 4 \\
&&&&&&&&&&&&&\\
\hline
&&&&&&&&&&&&&\\
0 & 40462 & -328405 & 1181862 & -2480440 & 3358633 & -3064605 & 1910758 & -808079 & 225280 & -39058 & 3728 & -133 & -3 \\
&&&&&&&&&&&&&\\
1 & 1089401 & -8312960 & 27967081 & -54487746 & 67891543 & -56371114 & 31518775 & -11720297 & 2794921 & -398308 & 29468 & -755 & -9 \\
&&&&&&&&&&&&&\\
2 & 15551542 & -111906456 & 352619477 & -638230462 & 731402730 & -551473014 & 275332506 & -89318075 & 17957805 & -2044330 & 109984 & -1701 & -6 \\
&&&&&&&&&&&&&\\
3 & 149809973 & -1022196296 & 3029937639 & -5111256667 & 5398506663 & -3699142263 & 1647523214 & -464542064 & 78061759 & -6958356 & 258399 & -2000 & -1 \\
&&&&&&&&&&&&&\\
4 & 1060193962 & -6903960379 & 19351278807 & -30549154463 & 29826756418 & -18606633971 & 7394683227 & -1808721923 & 252482556 & -17347693 & 424824 & -1365 & 0 \\
&&&&&&&&&&&&&\\
5 & 5749235605 & -35971215893 & 95853870247 & -142206408528 & 128744498873 & -73258030501 & 25991691895 & -5505943611 & 634915265 & -33123863 & 511071 & -560 & 0 \\
&&&&&&&&&&&&&\\
6 & 24515141237 & -148347082214 & 377772983884 & -528806354777 & 445198799642 & -231456580450 & 73327316027 & -13426393414 & 1271248262 & -49536205 & 458144 & -136 & 0 \\
&&&&&&&&&&&&&\\
7 & 83713528167 & -493054662213 & 1205735871217 & -1598297232698 & 1254433171038 & -596626214400 & 168735685696 & -26638174893 & 2056444657 & -58724268 & 307715 & -18 & 0 \\
&&&&&&&&&&&&&\\
8 & 232181721350 & -1339091523856 & 3158698278369 & -3977930045311 & 2916594182708 & -1270033416065 & 320393570085 & -43471409827 & 2714036622 & -55548598 & 154524 & -1 & 0 \\
&&&&&&&&&&&&&\\
9 & 529061326865 & -3005282202495 & 6865723701412 & -8237581076404 & 5651251963488 & -2253679329689 & 506430752402 & -58825334243 & 2942196810 & -42055645 & 57499 & 0 & 0 \\
&&&&&&&&&&&&&\\
10 & 999901634949 & -5625475790647 & 12492933041236 & -14314900382424 & 9199345356968 & -3358974796964 & 670974228930 & -66409836646 & 2632021622 & -25492601 & 15577 & 0 & 0 \\
&&&&&&&&&&&&&\\
11 & 1579912432814 & -8851638964768 & 19172932517151 & -21023714593994 & 12664940196393 & -4230717630716 & 749180958524 & -62830989756 & 1948414542 & -12343168 & 2978 & 0 & 0 \\
&&&&&&&&&&&&&\\
12 & 2100923882358 & -11784865427355 & 24973310181543 & -26250517882829 & 14827449083628 & -4525402328104 & 707888771163 & -49976382399 & 1194849756 & -4748141 & 380 & 0 & 0 \\
&&&&&&&&&&&&&\\
13 & 2364056295062 & -13348361985445 & 27751059919062 & -28002744326326 & 14828066667972 & -4126942351749 & 567743956597 & -33483312653 & 606575204 & -1437753 & 29 & 0 & 0 \\
&&&&&&&&&&&&&\\
14 & 2260885947021 & -12920004974340 & 26419687899169 & -25622229879786 & 12711183782080 & -3218120483560 & 387251430360 & -18907627955 & 254244735 & -337725 & 1 & 0 & 0 \\
&&&&&&&&&&&&&\\
15 & 1843899410104 & -10723805711918 & 21619800976560 & -20170802333741 & 9365038857728 & -2150060140301 & 224833996063 & -8992550361 & 87556031 & -60165 & 0 & 0 & 0 \\
&&&&&&&&&&&&&\\
16 & 1285500947246 & -7652697236144 & 15244035196288 & -13692589780602 & 5940325634334 & -1232087381017 & 111082353433 & -3594308673 & 24582979 & -7844 & 0 & 0 & 0 \\
&&&&&&&&&&&&&\\
17 & 767181318864 & -4703275369736 & 9275598325686 & -8026069091463 & 3246782576349 & -605653764269 & 46633085182 & -1202643219 & 5563311 & -705 & 0 & 0 & 0 \\
&&&&&&&&&&&&&\\
18 & 392093616730 & -2491467327515 & 4873731945238 & -4064415693849 & 1528956586250 & -255151779901 & 16586496221 & -334841196 & 998061 & -39 & 0 & 0 & 0 \\
&&&&&&&&&&&&&\\
19 & 171496884659 & -1137410904043 & 2210745244274 & -1777455042689 & 619655854309 & -91930972458 & 4975701525 & -76904083 & 138507 & -1 & 0 & 0 & 0 \\
&&&&&&&&&&&&&\\
20 & 64075193503 & -446974245545 & 864599171973 & -670349622143 & 215644648955 & -28231416371 & 1250647656 & -14392351 & 14323 & 0 & 0 & 0 & 0 \\
&&&&&&&&&&&&&\\
21 & 20384103423 & -150851963427 & 290831207393 & -217479954391 & 64210285686 & -7352495976 & 260973132 & -2156878 & 1038 & 0 & 0 & 0 & 0 \\
&&&&&&&&&&&&&\\
22 & 5494861982 & -43566210771 & 83829863705 & -60464623778 & 16274321473 & -1612596777 & 44636521 & -252402 & 47 & 0 & 0 & 0 & 0 \\
&&&&&&&&&&&&&\\
23 & 1246538879 & -10710673118 & 20596046283 & -14329011845 & 3485985544 & -295010570 & 6147029 & -22203 & 1 & 0 & 0 & 0 & 0 \\
&&&&&&&&&&&&&\\
24 & 235736043 & -2225506041 & 4281878288 & -2873287523 & 624945733 & -44429482 & 664362 & -1380 & 0 & 0 & 0 & 0 & 0 \\
&&&&&&&&&&&&&\\
25 & 36683212 & -387027796 & 745882601 & -482715579 & 92532599 & -5409211 & 54228 & -54 & 0 & 0 & 0 & 0 & 0 \\
&&&&&&&&&&&&&\\
26 & 4612971 & -55588795 & 107421034 & -67040066 & 11110494 & -518778 & 3141 & -1 & 0 & 0 & 0 & 0 & 0 \\
&&&&&&&&&&&&&\\
27 & 456837 & -6474484 & 12557437 & -7556246 & 1054047 & -37706 & 115 & 0 & 0 & 0 & 0 & 0 & 0 \\
&&&&&&&&&&&&&\\
28 & 34280 & -595786 & 1160837 & -673383 & 76001 & -1951 & 2 & 0 & 0 & 0 & 0 & 0 & 0 \\
&&&&&&&&&&&&&\\
29 & 1831 & -41666 & 81623 & -45636 & 3912 & -64 & 0 & 0 & 0 & 0 & 0 & 0 & 0 \\
&&&&&&&&&&&&&\\
30 & 62 & -2080 & 4100 & -2209 & 128 & -1 & 0 & 0 & 0 & 0 & 0 & 0 & 0 \\
&&&&&&&&&&&&&\\
31 & 1 & -66 & 131 & -68 & 2 & 0 & 0 & 0 & 0 & 0 & 0 & 0 & 0 \\
&&&&&&&&&&&&&\\
32 & 0 & -1 & 2 & -1 & 0 & 0 & 0 & 0 & 0 & 0 & 0 & 0 & 0 \\
&&&&&&&&&&&&&\\
\hline
\end{tabular}
}
\end{landscape}

\subsection{$SO/Sp$ Chern-Simons\label{No}}

In this case, the integers (\ref{icK1}) and (\ref{icK2}) could be calculated for not so many representations as compared with the HOMFLY case, since the HOMFLY polynomials in composite representations are not available so far for exception of the adjoint representation. The answers for this latter one can be obtained for the Kauffman and HOMFLY cases at once from the universal adjoint knot polynomials, \cite{MMkrM,MMuniv}. Their explicit expressions, as well as the Kauffman polynomials in the fundamental representation and the HOMFLY polynomials in the fundamental, first symmetric and first antisymmetric representations, which are also necessary in this case, can be found in \cite{knotebook}. The universal adjoint knot polynomials have been constructed so far only for the arborescent knots. Since knot $8_{20}$ used as an example in the previous subsection is arborescent, we give the integers for this knot (the results for other arborescent knots can be again found in \cite{knotebook}). In fact, knot $8_{20}$ enjoys a peculiar property: $\hat{\mathbf{N}}^{c=2}_{[ 1]}=0$ for it. This property was conjectured in \cite{Stevan} for the torus knots, however, it turns out to be the case for some other knots too, though it is not met too often: in the Rolfsen table \cite{katlas} only the knots $5_2, 7_1, 8_{20}, 9_1, 10_{125}, 10_{128}, 10_{132}, 10_{139}, 10_{161}$ and the torus knots $3_1,5_1,8_{19},10_{124}$ celebrate this property.

\bigskip

\textbf{Knot} $8_{20}$:

\bigskip

{\footnotesize
 \begin{tabular}{ccccc}
$\hat{\mathbf{N}}^{c=0}_{[ 1]}:$ &
\begin{tabular}{|c|cccc|}
\hline
&&&&\\
$ k \backslash n=$ & -5 & -3 & -1 & 1 \\
&&&&\\
\hline
&&&&\\
0 & 4 & -12 & 10 & -2 \\
&&&&\\
1 & 2 & -10 & 10 & -2 \\
&&&&\\
2 & 0 & -2 & 2 & 0 \\
&&&&\\
\hline
\end{tabular}
& $\hat{\mathbf{N}}^{c=1}_{[ 1]}:$ &
\begin{tabular}{|c|ccccc|}
\hline
&&&&&\\
$ k \backslash n=$ & -6 & -4 & -2 & 0 & 2 \\
&&&&&\\
\hline
&&&&&\\
0 & 3 & -10 & 12 & -5 & 1 \\
&&&&&\\
1 & 4 & -15 & 16 & -5 & 0 \\
&&&&&\\

2 & 1 & -7 & 7 & -1 & 0 \\
&&&&&\\
3 & 0 & -1 & 1 & 0 & 0 \\
&&&&&\\
\hline
\end{tabular}
& $\hat{\mathbf{N}}^{c=2}_{[ 1]}=0$
\end{tabular}

\begin{tabular}{cc}
$\hat{\mathbf{N}}^{c=0}_{[ 2]}:$ &
\begin{tabular}{|c|ccccccccc|}
\hline
&&&&&&&&&\\
$ k \backslash n=$ & -12 & -10 & -8 & -6 & -4 & -2 & 0 & 2 & 4 \\
&&&&&&&&&\\
\hline
&&&&&&&&&\\
0 & 9 & 22 & -216 & 462 & -442 & 194 & -25 & -6 & 1 \\
&&&&&&&&&\\
1 & 24 & 60 & -632 & 1340 & -1212 & 484 & -60 & -4 & 0 \\
&&&&&&&&&\\
2 & 22 & 84 & -817 & 1662 & -1364 & 458 & -45 & 0 & 0 \\
&&&&&&&&&\\
3 & 8 & 58 & -564 & 1106 & -804 & 208 & -12 & 0 & 0 \\
&&&&&&&&&\\
4 & 1 & 18 & -211 & 408 & -261 & 46 & -1 & 0 & 0 \\
&&&&&&&&&\\
5 & 0 & 2 & -40 & 78 & -44 & 4 & 0 & 0 & 0 \\
&&&&&&&&&\\
6 & 0 & 0 & -3 & 6 & -3 & 0 & 0 & 0 & 0 \\
&&&&&&&&&\\
\hline
\end{tabular}
\end{tabular}

\begin{tabular}{cc}
$\hat{\mathbf{N}}^{c=0}_{[ 1, 1]}:$ &
\begin{tabular}{|c|ccccccccc|}
\hline
&&&&&&&&&\\
$ k \backslash n=$ & -12 & -10 & -8 & -6 & -4 & -2 & 0 & 2 & 4 \\
&&&&&&&&&\\
\hline
&&&&&&&&&\\
0 & 9 & 40 & -300 & 620 & -594 & 272 & -45 & -4 & 1 \\
&&&&&&&&&\\
1 & 24 & 150 & -1050 & 2090 & -1864 & 762 & -110 & -2 & 0 \\
&&&&&&&&&\\
2 & 22 & 268 & -1685 & 3154 & -2524 & 854 & -89 & 0 & 0 \\
&&&&&&&&&\\
3 & 8 & 242 & -1500 & 2684 & -1904 & 498 & -28 & 0 & 0 \\
&&&&&&&&&\\
4 & 1 & 110 & -765 & 1352 & -857 & 162 & -3 & 0 & 0 \\
&&&&&&&&&\\
5 & 0 & 24 & -220 & 396 & -228 & 28 & 0 & 0 & 0 \\
&&&&&&&&&\\
6 & 0 & 2 & -33 & 62 & -33 & 2 & 0 & 0 & 0 \\
&&&&&&&&&\\
7 & 0 & 0 & -2 & 4 & -2 & 0 & 0 & 0 & 0 \\
&&&&&&&&&\\
\hline
\end{tabular}
\end{tabular}
}

\begin{landscape}
{\tiny
\begin{tabular}{cc}
$\hat{\mathbf{N}}^{c=1}_{[ 2]}:$ &
\begin{tabular}{|c|cccccccc|}
\hline
&&&&&&&&\\
$ k \backslash n=$ & -11 & -9 & -7 & -5 & -3 & -1 & 1 & 3 \\
&&&&&&&&\\
\hline
&&&&&&&&\\
0 & 163 & -723 & 1301 & -1217 & 631 & -179 & 25 & -1 \\
&&&&&&&&\\
1 & 1459 & -6030 & 9636 & -7506 & 2970 & -589 & 61 & -1 \\
&&&&&&&&\\
2 & 6463 & -25270 & 36235 & -23553 & 6911 & -836 & 50 & 0 \\
&&&&&&&&\\
3 & 17319 & -66196 & 86737 & -47418 & 10208 & -667 & 17 & 0 \\
&&&&&&&&\\
4 & 30172 & -116549 & 142053 & -65559 & 10204 & -323 & 2 & 0 \\
&&&&&&&&\\
5 & 35400 & -142768 & 164366 & -63878 & 6974 & -94 & 0 & 0 \\
&&&&&&&&\\
6 & 28479 & -123929 & 136524 & -44288 & 3229 & -15 & 0 & 0 \\
&&&&&&&&\\
7 & 15809 & -76877 & 81912 & -21832 & 989 & -1 & 0 & 0 \\
&&&&&&&&\\
8 & 6023 & -34068 & 35419 & -7565 & 191 & 0 & 0 & 0 \\
&&&&&&&&\\
9 & 1542 & -10670 & 10902 & -1795 & 21 & 0 & 0 & 0 \\
&&&&&&&&\\
10 & 253 & -2302 & 2325 & -277 & 1 & 0 & 0 & 0 \\
&&&&&&&&\\
11 & 24 & -325 & 326 & -25 & 0 & 0 & 0 & 0 \\
&&&&&&&&\\
12 & 1 & -27 & 27 & -1 & 0 & 0 & 0 & 0 \\
&&&&&&&&\\
13 & 0 & -1 & 1 & 0 & 0 & 0 & 0 & 0 \\
&&&&&&&&\\
\hline
\end{tabular}
\end{tabular}
\begin{tabular}{cc}
$\hat{\mathbf{N}}^{c=2}_{[ 2]}:$ &
\begin{tabular}{|c|ccccccccc|}
\hline
&&&&&&&&&\\
$ k \backslash n=$ & -12 & -10 & -8 & -6 & -4 & -2 & 0 & 2 & 4 \\
&&&&&&&&&\\
\hline
&&&&&&&&&\\
0 & 248 & -954 & 1413 & -1015 & 399 & -138 & 65 & -21 & 3 \\
&&&&&&&&&\\
1 & 2419 & -9185 & 12990 & -8445 & 2575 & -469 & 155 & -41 & 1 \\
&&&&&&&&&\\
2 & 10970 & -40936 & 54318 & -31193 & 7435 & -708 & 143 & -29 & 0 \\
&&&&&&&&&\\
3 & 28819 & -107691 & 134330 & -67226 & 12320 & -606 & 63 & -9 & 0 \\
&&&&&&&&&\\
4 & 47840 & -183768 & 217177 & -93757 & 12806 & -310 & 13 & -1 & 0 \\
&&&&&&&&&\\
5 & 52677 & -213996 & 241766 & -89041 & 8686 & -93 & 1 & 0 & 0 \\
&&&&&&&&&\\
6 & 39561 & -175117 & 190771 & -59093 & 3893 & -15 & 0 & 0 & 0 \\
&&&&&&&&&\\
7 & 20519 & -102241 & 108245 & -27663 & 1141 & -1 & 0 & 0 & 0 \\
&&&&&&&&&\\
8 & 7335 & -42733 & 44274 & -9086 & 210 & 0 & 0 & 0 & 0 \\
&&&&&&&&&\\
9 & 1772 & -12673 & 12926 & -2047 & 22 & 0 & 0 & 0 & 0 \\
&&&&&&&&&\\
10 & 276 & -2601 & 2625 & -301 & 1 & 0 & 0 & 0 & 0 \\
&&&&&&&&&\\
11 & 25 & -351 & 352 & -26 & 0 & 0 & 0 & 0 & 0 \\
&&&&&&&&&\\
12 & 1 & -28 & 28 & -1 & 0 & 0 & 0 & 0 & 0 \\
&&&&&&&&&\\
13 & 0 & -1 & 1 & 0 & 0 & 0 & 0 & 0 & 0 \\
&&&&&&&&&\\
\hline
\end{tabular}
\end{tabular}

\bigskip

\begin{tabular}{cc}
$\hat{\mathbf{N}}^{c=1}_{[ 1, 1]}:$ &
\begin{tabular}{|c|cccccccc|}
\hline
&&&&&&&&\\
$ k \backslash n=$ & -11 & -9 & -7 & -5 & -3 & -1 & 1 & 3 \\
&&&&&&&&\\
\hline
&&&&&&&&\\
0 & 208 & -944 & 1734 & -1642 & 842 & -222 & 24 & 0 \\
&&&&&&&&\\
1 & 2107 & -8865 & 14524 & -11641 & 4676 & -856 & 55 & 0 \\
&&&&&&&&\\
2 & 10561 & -41834 & 61612 & -41798 & 12948 & -1529 & 40 & 0 \\
&&&&&&&&\\
3 & 32160 & -123599 & 166318 & -95920 & 22662 & -1632 & 11 & 0 \\
&&&&&&&&\\
4 & 64264 & -247000 & 308745 & -151799 & 26894 & -1105 & 1 & 0 \\
&&&&&&&&\\
5 & 87697 & -347013 & 408847 & -171220 & 22160 & -471 & 0 & 0 \\
&&&&&&&&\\
6 & 83551 & -350248 & 393823 & -139736 & 12731 & -121 & 0 & 0 \\
&&&&&&&&\\
7 & 56187 & -256993 & 278670 & -82900 & 5053 & -17 & 0 & 0 \\
&&&&&&&&\\
8 & 26713 & -137658 & 145204 & -35610 & 1352 & -1 & 0 & 0 \\
&&&&&&&&\\
9 & 8897 & -53635 & 55429 & -10923 & 232 & 0 & 0 & 0 \\
&&&&&&&&\\
10 & 2026 & -14998 & 15275 & -2326 & 23 & 0 & 0 & 0 \\
&&&&&&&&\\
11 & 300 & -2927 & 2952 & -326 & 1 & 0 & 0 & 0 \\
&&&&&&&&\\
12 & 26 & -378 & 379 & -27 & 0 & 0 & 0 & 0 \\
&&&&&&&&\\
13 & 1 & -29 & 29 & -1 & 0 & 0 & 0 & 0 \\
&&&&&&&&\\
14 & 0 & -1 & 1 & 0 & 0 & 0 & 0 & 0 \\
&&&&&&&&\\
\hline
\end{tabular}
\end{tabular}
 \begin{tabular}{cc}
$\hat{\mathbf{N}}^{c=2}_{[ 1, 1]}:$ &
\begin{tabular}{|c|ccccccccc|}
\hline
&&&&&&&&&\\
$ k \backslash n=$ & -12 & -10 & -8 & -6 & -4 & -2 & 0 & 2 & 4 \\
&&&&&&&&&\\
\hline
&&&&&&&&&\\
0 & 315 & -1214 & 1794 & -1260 & 438 & -100 & 36 & -10 & 1 \\
&&&&&&&&&\\
1 & 3465 & -13225 & 18885 & -12420 & 3714 & -470 & 66 & -15 & 0 \\
&&&&&&&&&\\
2 & 17878 & -67003 & 90314 & -53437 & 13273 & -1060 & 42 & -7 & 0 \\
&&&&&&&&&\\
3 & 53910 & -201207 & 255638 & -133587 & 26585 & -1349 & 11 & -1 & 0 \\
&&&&&&&&&\\
4 & 103753 & -394200 & 474873 & -216810 & 33397 & -1014 & 1 & 0 & 0 \\
&&&&&&&&&\\
5 & 134083 & -531391 & 611656 & -241661 & 27769 & -456 & 0 & 0 & 0 \\
&&&&&&&&&\\
6 & 120036 & -509029 & 564214 & -190755 & 15654 & -120 & 0 & 0 & 0 \\
&&&&&&&&&\\
7 & 75736 & -353077 & 379598 & -108244 & 6004 & -17 & 0 & 0 & 0 \\
&&&&&&&&&\\
8 & 33858 & -178831 & 187707 & -44274 & 1541 & -1 & 0 & 0 & 0 \\
&&&&&&&&&\\
9 & 10648 & -66054 & 68079 & -12926 & 253 & 0 & 0 & 0 & 0 \\
&&&&&&&&&\\
10 & 2301 & -17575 & 17875 & -2625 & 24 & 0 & 0 & 0 & 0 \\
&&&&&&&&&\\
11 & 325 & -3277 & 3303 & -352 & 1 & 0 & 0 & 0 & 0 \\
&&&&&&&&&\\
12 & 27 & -406 & 407 & -28 & 0 & 0 & 0 & 0 & 0 \\
&&&&&&&&&\\
13 & 1 & -30 & 30 & -1 & 0 & 0 & 0 & 0 & 0 \\
&&&&&&&&&\\
14 & 0 & -1 & 1 & 0 & 0 & 0 & 0 & 0 & 0 \\
&&&&&&&&&\\
\hline
\end{tabular}
\end{tabular}
}
\end{landscape}

\subsection{Link polynomials\label{Nl}}

In the link case, already the lowest relations (\ref{hL}), (\ref{gL}) imply a non-trivial test: one has to check that the expansions (\ref{icKl1}) and (\ref{icKl2}), indeed, starts from 1 and $(q-q^{-1})$ respectively, i.e. that (\ref{gL}) cancels at $q=1$, while (\ref{hL}) is regular. The literal integrality checks in this case require knowledge of a series of colored knot and link polynomials. We need to know: the HOMFLY and Kauffman polynomials for links and knots in the fundamental representation, which can be found in \cite{katlas}; the HOMFLY polynomials of links when one of the link components is in the first (anti)symmetric representation, and the other one is in the fundamental one, which are calculated using the known exclusive Racah matrices \cite{RZ} or by the cabling method \cite{RTmod2} and can be found in \cite{knotebook}; the HOMFLY polynomials of knots in the adjoint representation and the Kauffman polynomials of knots in the first (anti)symmetric representation, which are read off the universal knot polynomials \cite{MMuniv} in \cite{knotebook}; the HOMFLY polynomials of links with one component in the adjoint representation and the other one in the fundamental one and similarly the Kauffman polynomials of links with one component in the first (anti)symmetric representation and the other one in the fundamental one, which can be constructed with the inclusive Racah matrices that we discuss in the next subsection (the manifest expressions for knot polynomials can be found in \cite{knotebook}). This finally allows us to obtain the integers, and we again write down them just for a link 7a3, while more examples can be found in \cite{knotebook}:

\bigskip

{\footnotesize
\hspace{-2.3cm} \begin{tabular}{cccccc}
$\hat{\mathbf{N}}^{c=0}_{[ 1],[ 1]}:$ &
\begin{tabular}{|c|cccc|}
\hline
&&&&\\
$ k \backslash n=$ & 0 & 2 & 4 & 6 \\
&&&&\\
\hline
&&&&\\
0 & 8 & -24 & 24 & -8 \\
&&&&\\
1 & 4 & -20 & 20 & -4 \\
&&&&\\
2 & 0 & -4 & 4 & 0 \\
&&&&\\
\hline
\end{tabular}
&
$\hat{\mathbf{N}}^{c=0}_{[ 2],[ 1]}:$ &
\begin{tabular}{|c|ccccc|}
\hline
&&&&&\\
$ k \backslash n=$ & -1 & 1 & 3 & 5 & 7 \\
&&&&&\\
\hline
&&&&&\\
0 & -4 & 24 & -48 & 40 & -12 \\
&&&&&\\
1 & -2 & 16 & -40 & 32 & -6 \\
&&&&&\\
2 & 0 & 2 & -8 & 6 & 0 \\
&&&&&\\
\hline
\end{tabular}
&
$\hat{\mathbf{N}}^{c=0}_{[ 1, 1],[ 1]}:$ &
\begin{tabular}{|c|ccccc|}
\hline
&&&&&\\
$ k \backslash n=$ & -1 & 1 & 3 & 5 & 7 \\
&&&&&\\
\hline
&&&&&\\
0 & -12 & 40 & -48 & 24 & -4 \\
&&&&&\\
1 & -6 & 32 & -40 & 16 & -2 \\
&&&&&\\
2 & 0 & 6 & -8 & 2 & 0 \\
&&&&&\\
\hline
\end{tabular}
\end{tabular}

\bigskip

\hspace{-1cm} \begin{tabular}{cccccc}
$\hat{\mathbf{N}}^{c=1}_{[ 1],[ 1]}:$ &
\begin{tabular}{|c|ccccc|}
\hline
&&&&&\\
$ k \backslash n=$ & 0 & 2 & 4 & 6 & 8 \\
&&&&&\\
\hline
&&&&&\\
0 & 0 & 0 & 0 & 0 & 0 \\
&&&&&\\
1 & 2 & -5 & 3 & 1 & -1 \\
&&&&&\\
2 & 1 & -5 & 4 & 0 & 0 \\
&&&&&\\
3 & 0 & -1 & 1 & 0 & 0 \\
&&&&&\\
\hline
\end{tabular}
&
$\hat{\mathbf{N}}^{c=1}_{[ 2],[ 1]}:$ &
\begin{tabular}{|c|cccc|}
\hline
&&&&\\
$ k \backslash n=$ & 0 & 2 & 4 & 6 \\
&&&&\\
\hline
&&&&\\
0 & 0 & 0 & 0 & 0 \\
&&&&\\
1 & 7 & -9 & 5 & -1 \\
&&&&\\
2 & 6 & -9 & 5 & -1 \\
&&&&\\
3 & 1 & -2 & 1 & 0 \\
&&&&\\
\hline
\end{tabular}
&
$\hat{\mathbf{N}}^{c=1}_{[ 1, 1],[ 1]}:$ &
\begin{tabular}{|c|cccc|}
\hline
&&&&\\
$ k \backslash n=$ & 0 & 2 & 4 & 6 \\
&&&&\\
\hline
&&&&\\
0 & 0 & 0 & 0 & 0 \\
&&&&\\
1 & 21 & -15 & -1 & 3 \\
&&&&\\
2 & 25 & -20 & 0 & 1 \\
&&&&\\
3 & 9 & -8 & 0 & 0 \\
&&&&\\
4 & 1 & -1 & 0 & 0 \\
&&&&\\
\hline
\end{tabular}
\end{tabular}}

\bigskip

{\tiny
\hspace{-2cm} \begin{tabular}{cccccc}
$\hat{\mathbf{N}}^{c=2}_{[1],[ 1]}:$ &
\begin{tabular}{|c|ccccc|}
\hline
&&&&&\\
$ k \backslash n=$ & 1 & 3 & 5 & 7 & 9 \\
&&&&&\\
\hline
&&&&&\\
0 & -5 & 14 & -12 & 2 & 1 \\
&&&&&\\
1 & -2 & 11 & -10 & 1 & 0 \\
&&&&&\\
2 & 0 & 2 & -2 & 0 & 0 \\
&&&&&\\
\hline
\end{tabular}
&
$\hat{\mathbf{N}}^{c=2}_{[ 2],[ 1]}:$ &
\begin{tabular}{|c|cccccc|}
\hline
&&&&&&\\
$ k \backslash n=$ & -1 & 1 & 3 & 5 & 7 & 9 \\
&&&&&&\\
\hline
&&&&&&\\
0 & 2 & -15 & 30 & -20 & 0 & 3 \\
&&&&&&\\
1 & 1 & -11 & 31 & -22 & 0 & 1 \\
&&&&&&\\
2 & 0 & -2 & 10 & -8 & 0 & 0 \\
&&&&&&\\
3 & 0 & 0 & 1 & -1 & 0 & 0 \\
&&&&&&\\
\hline
\end{tabular}
&
$\hat{\mathbf{N}}^{c=2}_{[ 1, 1],[ 1]}:$ &
\begin{tabular}{|c|cccccc|}
\hline
&&&&&&\\
$ k \backslash n=$ & -1 & 1 & 3 & 5 & 7 & 9 \\
&&&&&&\\
\hline
&&&&&&\\
0 & 6 & -21 & 26 & -12 & 0 & 1 \\
&&&&&&\\
1 & 5 & -22 & 27 & -10 & 0 & 0 \\
&&&&&&\\
2 & 1 & -8 & 9 & -2 & 0 & 0 \\
&&&&&&\\
3 & 0 & -1 & 1 & 0 & 0 & 0 \\
&&&&&&\\
\hline
\end{tabular}
\end{tabular}
}

\subsection{Racah matrices for links}

In this subsection, we write down the inclusive Racah matrices that are necessary in order to perform calculations for checking the integrality conjectures in the case of 2-component links in the previous subsection. Non-trivial (new) Racah matrices are required in the following cases:

\paragraph{HOMFLY polynomials with one component in the adjoint representation and the other one in the fundamental one.} In this case, one studies the product of $SU(N)$ representations
\be
[1]\otimes[1]\otimes ([1],[1])=([3],[1])+2([2,1],[1])+2[2]+2[1,1]+([1,1,1],[1])
\ee
and one can use the eigenvalue conjecture for links, \cite{RTmod2} in order to construct the inclusive 2x2 matrices. Note that, since we are dealing here with links, there are two different matrices for each representation \cite{RTmod2}. The eigenvalues
(diagonalized $\mathcal{R}$-matrices, \cite{MMMkn12}) are
\be
\begin{array}{l}
{\mathcal{R}}_{([3],[1]);xx}=\left(Aq^{-1}\right),\ \ \ \ {\mathcal{R}}_{([1,1,1],[1]);xx}=\left(-Aq\right),\ \ \ \ {\mathcal{R}}_{([3],[1]);xy}={\mathcal{R}}_{([1,1,1],[1]);xy}=\left(q\right)
\\
{\mathcal{R}}_{([2,1],[1]);xx}=\left(\begin{array}{cc}Aq^{-1}&\\&-Aq\end{array}\right),\ \ \ \ {\mathcal{R}}_{([2,1],[1]);xy}=\left(\begin{array}{cc}q&\\&-q^{-1}\end{array}\right),
\\
{\mathcal{R}}_{[2];xx}=\left(\begin{array}{cc}Aq^{-1}&\\&-Aq\end{array}\right),\ \ \ \ {\mathcal{R}}_{[2];xy}=\left(\begin{array}{cc}A&\\&-q^{-1}\end{array}\right),
\\
{\mathcal{R}}_{[1,1];xx}=\left(\begin{array}{cc}Aq^{-1}&\\&-Aq\end{array}\right),\ \ \ \ {\mathcal{R}}_{[1,1];xy}=\left(\begin{array}{cc}A&\\&q\end{array}\right)
\end{array}
\ee
and the mixing (inclusive Racah) matrices are
\be
\begin{array}{l}
U_{([3],[1]);xxy}=U_{([1,1,1],[1]);xxy}=U_{([3],[1]);xyx}=U_{([1,1,1],[1]);xyx}=\left(1\right)
\\
U_{([2,1],[1]);xxy}=\left(\begin{array}{cc}-\frac{1}{[2]}&\frac{\sqrt{[3]}}{[2]}\\ \frac{\sqrt{[3]}}{[2]}&\frac{1}{[2]}\end{array}\right),\ \ \ \ U_{([2,1],[1]);xyx}=\left(\begin{array}{cc}-\frac{1}{[2]}&-\frac{\sqrt{[3]}}{[2]}\\-\frac{\sqrt{[3]}}{[2]}&\frac{1}{[2]}\end{array}\right),
\\
U_{[2];xxy}=\left(\begin{array}{cc}-\sqrt{\frac{D_0}{D_1[2]}}&\sqrt{\frac{D_2}{D_1[2]}}\\\sqrt{\frac{D_2}{D_1[2]}}&\sqrt{\frac{D_0}{D_1[2]}}\end{array}\right),\ \ \ \ U_{[2];xyx}=\left(\begin{array}{cc}-\frac{1}{D_1}&\frac{\sqrt{D_2D_0}}{D_{1}}\\\frac{\sqrt{D_2D_0}}{D_{1}}&\frac{1}{D_1}\end{array}\right),
\\
U_{[1,1];xxy}=\left(\begin{array}{cc}-\sqrt{\frac{D_{-2}}{D_{-1}[2]}}&\sqrt{\frac{D_0}{D_{-1}[2]}}\\\sqrt{\frac{D_0}{D_{-1}[2]}}&\sqrt{\frac{D_{-2}}{D_{-1}[2]}}\end{array}\right),\ \ \ \ U_{[1,1];xyx}=\left(\begin{array}{cc}-\frac{1}{D_{-1}}&\frac{\sqrt{D_{-2}D_0}}{D_{-1}}\\\frac{\sqrt{D_{-2}D_0}}{D_{-1}}&\frac{1}{D_{-1}}\end{array}\right)
\end{array}
\ee
where $D_i=(Aq^i-A^{-1}q^{-i})/(q-q^{-1})$.

\paragraph{Kauffman polynomials in the fundamental representation.}
In this case (when the answers can be also found in \cite{katlas}), one studies the product of $SO(N)$ representations
\be
[1]\otimes[1]\otimes [1]=[3]+2[2,1]+[1,1,1]+3[1]
\ee
and both the 2x2 and 3x3 inclusive Racah matrices (since all three representations are the same, there is only one matrix for each representation) can be obtained from the eigenvalue conjecture for knots, \cite{IMMMev}. The eigenvalues are
\be
{\mathcal{R}}_{[3]}=(Aq),\ \ \ \ \ \ \ {\mathcal{R}}_{[1,1,1]}=(-A/q),\ \ \ \ \ \ \
{\mathcal{R}}_{[2,1]}=\left(\begin{array}{cc}Aq&\\&-A/q\end{array}\right),\ \ \ \ \ \ \
{\mathcal{R}}_{[1]}=\left(\begin{array}{ccc}Aq&&\\&-A/q&\\&&-A^2\end{array}\right)
\ee
and, hence, the Racah matrices are
\be
U_{[3]}=U_{[1,1,1]}=(1)\\
U_{[2,1]}={1\over [2]_q}\left(\begin{array}{cc}
1&\sqrt{[3]_q}\\
\sqrt{[3]_q}&-1
\end{array}\right)
\ee
\be
U_{[1]}={1\over [2]_q}\left(\begin{array}{ccc}
\displaystyle{{Aq+1\over A+q}}&\displaystyle{-{1\over q}\sqrt{(Aq^3-1)(A+q^3)\over (Aq-1)(A+q)}}&\displaystyle{\sqrt{[2]_q{(A^2-1)(A+q^3)\over (Aq-1)(A+q)}}}\\
-\displaystyle{{1\over q}\sqrt{(Aq^3-1)(A+q^3)\over (Aq-1)(A+q)}}&\displaystyle{{A-q\over Aq-1}}&\displaystyle{\sqrt{[2]_q{(A^2-1)(Aq^3-1)\over (Aq-1)(A+q)}}}\\
\displaystyle{\sqrt{[2]_q{(A^2-1)(A+q^3)\over (Aq-1)(A+q)}}}&\displaystyle{\sqrt{[2]_q{(A^2-1)(Aq^3-1)\over (Aq-1)(A+q)}}}&[2]_qA\displaystyle{{q^2-1\over (Aq-1)(A+q)}}
\end{array}\right)
\ee
where $[n]_q=(q^n-q^{-n})/(q-q^{-1})$ denotes the usual quantum number.

\paragraph{Kauffman polynomials with one component in the (anti)symmetric representation and the other one in the fundamental one.}
In this case, one studies the product of $SO(N)$ representations
\be
[1]\otimes[1]\otimes [2]=[4]+2[3,1]+[2,2]+[2,1,1]+3[2]+2[1,1]+1
\ee
The 2x2 inclusive Racah matrices can be again read off the eigenvalue conjecture, the eigenvalues being
\be
\begin{array}{l}
{\mathcal{R}}_{[4];xx}={\mathcal{R}}_{[2,2];xx}={\mathcal{R}}_{[\emptyset];xx}=\left(Aq^{-1}\right),\ \ \ \ {\mathcal{R}}_{[2,1,1];xx}=\left(-Aq\right),\\ {\mathcal{R}}_{[4];xy}=\left(-q^{-2}\right),\ \ \ \ {\mathcal{R}}_{[2,2];xy}={\mathcal{R}}_{[2,1,1];xy}=\left(q\right),\ \ \ \ {\mathcal{R}}_{[\emptyset];xy}=\left(-Aq\right)
\\
{\mathcal{R}}_{[3,1];xx}=\left(\begin{array}{cc}Aq^{-1}&\\&-Aq\end{array}\right),\ \ \ \ {\mathcal{R}}_{[3,1];xy}=\left(\begin{array}{cc}-q^{-2}&\\&-q\end{array}\right),
\\
{\mathcal{R}}_{[1,1];xx}=\left(\begin{array}{cc}Aq^{-1}&\\&-Aq\end{array}\right),\ \ \ \ {\mathcal{R}}_{[1,1];xy}=\left(\begin{array}{cc}q&\\&-Aq\end{array}\right)
\\
{\mathcal{R}}_{([2]);xx}=\left(\begin{array}{ccc}Aq^{-1}&&\\&-Aq&\\&&A^2\end{array}\right),\ \ \ \ {\mathcal{R}}_{([2]);xy}=\left(\begin{array}{ccc}q^{-2}&&\\&-q&\\&&Aq\end{array}\right),
\end{array}
\ee
and the mixing matrices being
\be
\begin{array}{l}
U_{[3,1];xxy}=\left(\begin{array}{cc}-\frac{1}{\sqrt{[3]}}&\sqrt{\frac{[4]}{[2][3]}}\\ \sqrt{\frac{[4]}{[2][3]}}&\frac{1}{\sqrt{[3]}}\end{array}\right),\ \ \ \ U_{[3,1];xyx}=\left(\begin{array}{cc}-\frac{1}{\sqrt{[3]}}&-\frac{\sqrt{[2][4]}}{[3]}\\-\frac{\sqrt{[2][4]}}{[3]}&\frac{1}{\sqrt{[3]}}\end{array}\right),
\\
U_{[1,1];xxy}=\left(\begin{array}{cc}-\sqrt{\frac{D_{-1}}{D_{0}[2]}}&\sqrt{\frac{D_{1}}{D_{0}[2]}}\\\sqrt{\frac{D_{1}}{D_{0}[2]}}&\sqrt{\frac{D_{-1}}{D_{0}[2]}}\end{array}\right),\ \ \ \ U_{[1,1];xyx}=\left(\begin{array}{cc}-\frac{1}{D_{0}}&-\frac{\sqrt{D_1D_{-1}}}{D_{0}}\\-\frac{\sqrt{D_1D_{-1}}}{D_{0}}
&\frac{1}{D_{0}}\end{array}\right)
\end{array}
\ee
while the 3x3 matrices\footnote{
Due to the relations between $SO(N)$ and $SU(N)$ theories particular cases of these matrices can be used for calculations of HOMFLY polynomials. Namely, due to similarity between $SU(2)$ and $SO(3)$ groups by substituting $A=q^4$ and $q=q^2$ one gets matrix $3\times 3$ for representation $[6,2]$ from the tensor product $[2]\otimes[2]\otimes[4]$. And due to similarity between $SO(6)$ and $SU(4)$ by substituting $A=q^5$ one gets matrix $3\times 3$ for representation $[3,3,1,1]$ from the tensor product $[1,1]\otimes[1,1]\otimes[2,2]$.
}
are of the form
\be
\begin{array}{l}
U_{[2];xxy}= \left(\begin{array}{ccc}
\sqrt{\frac{(A-q)q}{(Aq^3-1)[3]}} &
\sqrt{(A+q^3)(Aq^5-1)}{Aq^4(A-A^{-1})[2][3]} &
\sqrt{\frac{(Aq^5-1)(Aq+1)(A-q)}{Aq^2(Aq^3-1)(A-A^{-1})[2]}} \\
\sqrt{\frac{(Aq^5-1)(A+q^3)}{q^2(Aq^3-1)(A+q)[3]}} &
\sqrt{\frac{(Aq^3+1)^2(A-q)}{q^3(A+q)(A^2-1)[2][3]}}
&-\sqrt{\frac{(A^2q^2-1)(Aq-1)(A+q^3)}{q(Aq^3-1)(A^2-1)(A+q)[2]}} \\
\sqrt{\frac{(Aq^5-1)(Aq+1)}{q(Aq^3-1)(A+q)[3]}} &
-\sqrt{\frac{(Aq+1)(A+q^3)(A-q)[2]}{q^2(A^2-1)(A+q)[3]}} &
\sqrt{\frac{A^2[2](q^2-1)^2}{(Aq^3-1)(A+q)(A^2-1)}}\end{array}\right)
\\
U_{[2];xyx}= \left(\begin{array}{ccc}
\frac{q(A-q)}{(Aq^3-1)[3]} &
\sqrt{\frac{(A+q^3)(A-q)(Aq^5-1)[2]}{q^3(Aq^3-1)(A^2-1)[3]^2}} &
\sqrt{\frac{(Aq-1)(Aq^5-1)(A^2q^2-1)[2]}{q(Aq^3-1)^2(A^2-1)[3]}} \\
\sqrt{\frac{(A+q^3)(A-q)(Aq^5-1)[2]}{q^3(Aq^3-1)(A^2-1)[3]^2}} &
\frac{A^2q^4+A^2q^2-Aq^3+Aq-q^2-1}{q^2(A^2-1)[3]} &
-\sqrt{\frac{(A-q)(Aq-1)(A^2q^2-1)(A+q^3)}{q^2(Aq^3-1)(A^2-1)^2[3]}} \\
\sqrt{\frac{(Aq-1)(Aq^5-1)(A^2q^2-1)[2]}{q(Aq^3-1)^2(A^2-1)[3]}} &
-\sqrt{\frac{(A-q)(Aq-1)(A^2q^2-1)(A+q^3)}{q^2(Aq^3-1)(A^2-1)^2[3]}} &
\frac{A(A-q)(q^2-1)}{(Aq^3-1)(A^2-1)}\end{array}\right),
\end{array}
\ee

\section{Gaussian distributions of the LMOV numbers $N$\label{gauss}}

In this section, we make clear a specific behaviour of the LMOV numbers that can be understood from analyses of results of Sections \ref{intests} and \ref{No}.
Namely, we plot the numbers for the knot $8_{20}$ from our tables in Sections \ref{intests} and \ref{No} against $k$. From these pictures, one can easily see that these numbers have Gaussian distributions(!). One can also see that the difference with the Gaussian distribution decreases with the size of the representation (since in the $SO/Sp$ case only the first symmetric representation is available, the accuracy is lower in this case). This is a simplest corollary of the results obtained in this paper, its theoretical meaning and implications of this spectacular fact as well as other applications of the results obtained in this paper will be discussed elsewhere.

\begin{figure}[h!]
\centering\leavevmode
\includegraphics[width=6cm]{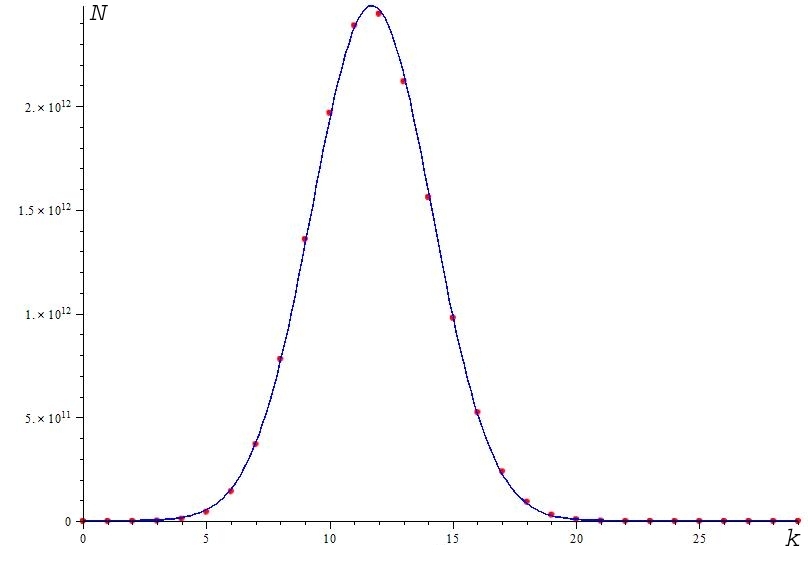}
\caption{Red dots are $N_{[3,1],-16,k}$ numbers, blue curve is a Gaussian distribution with parameters $\mu=11.7,\ \sigma=2.42,\ I=1.51\cdot10^{13}$}
\label{gauss2}
\end{figure}

\begin{figure}[h!]
\centering\leavevmode
\includegraphics[width=6cm]{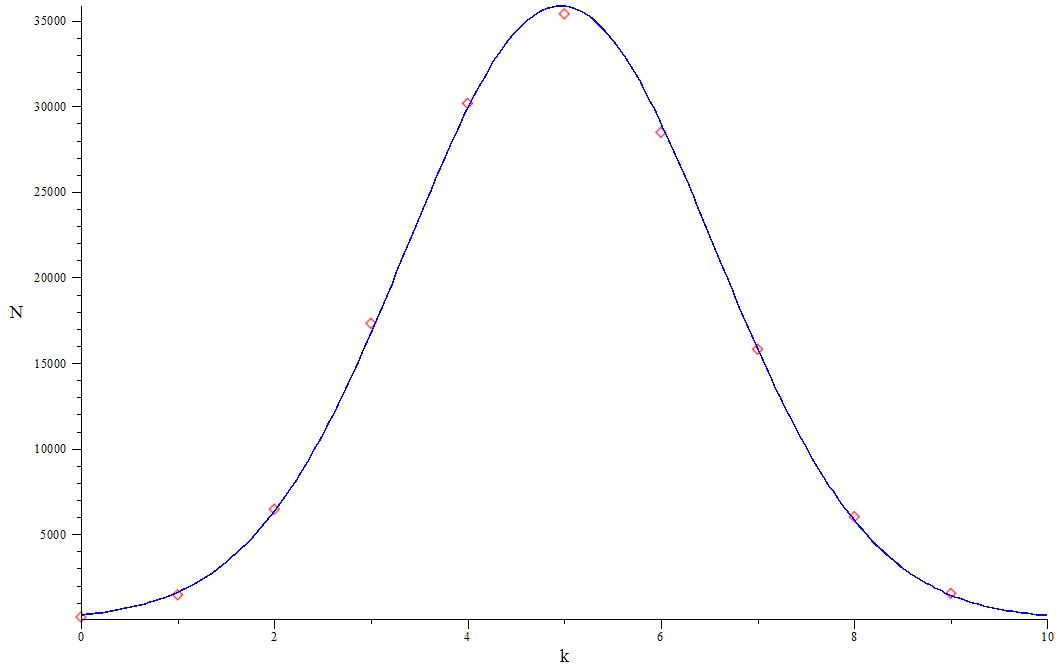}
\caption{Red dots are $\hat{N}^{c=1}_{[2],-11,k}$ numbers, blue curve is a Gaussian distribution with parameters $\mu=4.96$  $\sigma=1.59$ $I=1.43\cdot 10^5$}
\label{gauss3}
\end{figure}

\begin{figure}[h!]
\centering\leavevmode
\includegraphics[width=6cm]{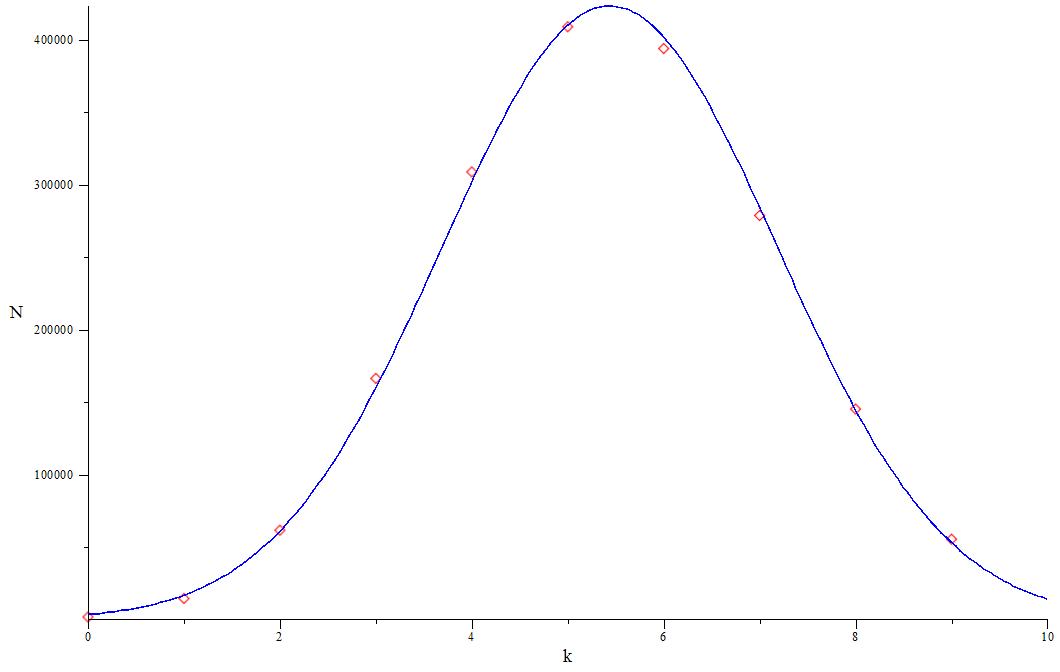}
\caption{Red dots are $\hat{N}^{c=1}_{[1,1],-7,k}$ numbers, blue curve is a Gaussian distribution with parameters $\mu=5.44$  $\sigma=1.75$ $I=1.85\ 10^6$}
\label{gauss4}
\end{figure}

Thus, it turns out that with a very good accuracy for any given knot ${\cal K}$, representation $Q$ and parameter $n$, the LMOV numbers as a function of
remaining variable $k$ are described by the formula
\be
\boxed{
N^{\cal K}_{Q,n,k} \approx
G^{\cal K}_{Q,n}(k) \equiv
\frac{I_{Q,n}^{\cal K} }{\sqrt{2\pi}\,\sigma_{Q,n}^{\cal K} }
\exp\left( -\frac{(k-\mu_{Q,n}^{\cal K} )^2}{2\sigma_{Q,n}^2}\right)
}
\ee
with only three parameters $\mu$, $\sigma$ and $I$ which can depend on ${\cal K}$, $Q$ and $n$, but not on $k$. Accuracy of this formula is illustrated in Figures \ref{gauss1} and \ref{gauss2} for the LMOV invariants in the $SU(N)$ case and \ref{gauss3}, \ref{gauss4} for the LMOV invariants in the $SO/Sp$ case:

\begin{figure}[h!]
\centering\leavevmode
\includegraphics[width=6cm]{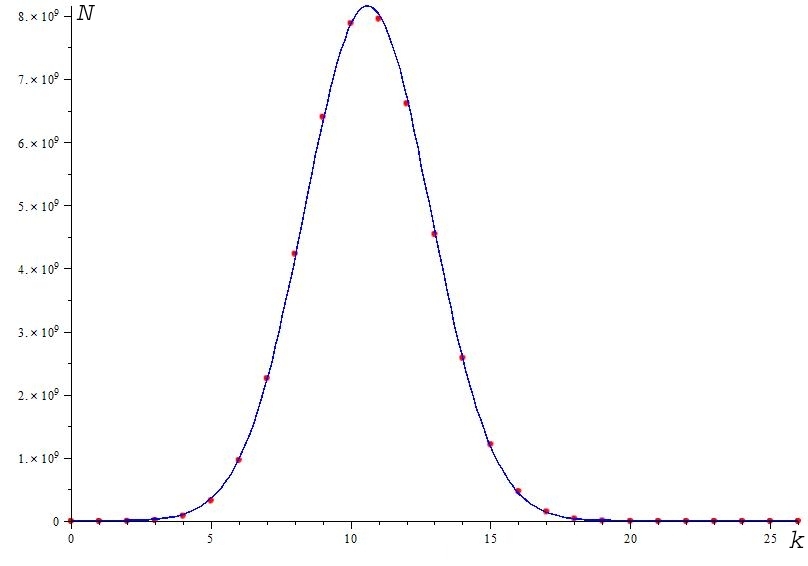}
\caption{Red dots are $N_{[4],-20,k}$ numbers, blue curve is a Gaussian distribution with parameters $\mu=10.6,\ \sigma=2.24,\ I=4.56 \cdot 10^{10}$}
\label{gauss1}
\end{figure}

\section{Conclusion
\label{conc}}

In this paper, we reported a positive result of new tests of the LMOV
integrality conjectures, made possible by a recent progress
in evaluation of the colored knot polynomials and first corollary of the performed calculations: the Gaussian distribution of the LMOV invariants.
The progress is a cumulative effect of merging of the different
research directions:

\begin{itemize}
\item[i)] reformulation of the RT formalism in the spaces of intertwining operators \cite{MMMkn12}-\cite{RTmod2}
with developments of the highest weight technique \cite{MMMS21,MMMS31} and
the eigenvalue conjecture \cite{IMMMev,MMuniv} to evaluate the inclusive and exclusive
Racah matrices \cite{MMMS31}-\cite{MMMSsu},

\item[ii)] representing the knot polynomials for all arborescent knots
through the exclusive Racah matrices $S$ and $\bar S$ \cite{MMMRV},

\item[iii)] developing the family technique \cite{MMMevo,MMfam,MMMRSS} in order to
adequately classify knots, at least, for calculational purposes,

\item[iv)] applying Vogel's universality \cite{Vog} to handle the
adjoint representations and their descendants; important here is that
deviations from the universality at the group theory level are not seen
in knot polynomial calculus \cite{MMkrM,MMuniv}.
\end{itemize}

The work in all these directions is hard, but interesting and important (see also a new development in another direction of an integrality conjecture for superpolynomials, \cite{Nawata}).
The integrality tests are a non-trivial application of its results,
and provide an additional stimulus for new advances.

\section*{Acknowledgements}

Our work is partly supported by RFBR grants 16-01-00291 (A.Mir.), 16-02-01021 (A.Mor.), mol-a-dk 16-32-60047 (An.Mor), mol-a-dk 16-31-60082 and MK-8769.2016.1 (A.S.) and by joint grants 17-51-50051-YaF, 15-51-52031-NSC-a, 16-51-45029-Ind-a and 16-51-53034-GFEN. PR, VKS and SD acknowledge DST-RFBR grant (INT/RUS/RFBR/P-231) for support. SD would like to thank CSIR for research fellowship.

\end{document}